\documentclass[useAMS,usenatbib]{mn2e}

\usepackage{graphicx, bm, amssymb, xcolor}
\usepackage{verbatim}
\usepackage[breaklinks,colorlinks,citecolor=blue]{hyperref}
\usepackage[all]{hypcap}

\usepackage{epsfig}
\usepackage{aas_macros}
\usepackage{natbib}
\usepackage{times}

\newbox\grsign \setbox\grsign=\hbox{$>$} \newdimen\grdimen \grdimen=\ht\grsign
\newbox\labox \newbox\gabox \newbox\simpropbox \newbox\wtildebox 
\setbox\gabox=\hbox{\raise.5ex\hbox{$>$}\llap
    {\lower.5ex\hbox{$\sim$}}}\ht1=\grdimen\dp1=0pt
\setbox\labox=\hbox{\raise.5ex\hbox{$<$}\llap
    {\lower.5ex\hbox{$\sim$}}}\ht2=\grdimen\dp2=0pt
\def\ga{\mathrel{\copy\gabox}}
\def\la{\mathrel{\copy\labox}}
\newcommand{\msun}{\mbox{M$_\odot$}}
\newcommand{\yr}{\mbox{${\rm yr}$}}
\newcommand{\myr}{\mbox{${\rm Myr}$}}
\newcommand{\gyr}{\mbox{${\rm Gyr}$}}
\newcommand{\pc}{\mbox{${\rm pc}$}}
\newcommand{\kpc}{\mbox{${\rm kpc}$}}
\newcommand{\kms}{\mbox{${\rm km}~{\rm s}^{-1}$}}
\newcommand{\cmc}{\mbox{${\rm cm}^{-3}$}}
\newcommand{\kcmc}{\mbox{${\rm K}~{\rm cm}^{-3}$}}
\newcommand{\feh}{\mbox{$[{\rm Fe}/{\rm H}]$}}

\newcommand{\be}{\begin{equation}}
\newcommand{\ee}{\end{equation}}
\newcommand{\bea}{\begin{eqnarray}}
\newcommand{\eea}{\end{eqnarray}}

\title{Globular clusters as the relics of regular star formation in `normal' high-redshift galaxies}
\author{J.~M.~Diederik Kruijssen\\
Max-Planck Institut f\"{u}r Astrophysik, Karl-Schwarzschild-Stra\ss e 1, 85748 Garching, Germany; kruijssen@mpa-garching.mpg.de}

\begin{document}

\date{Accepted 2015 September 1. Received 2015 August 27; in original form 2014 September 29.}

\pagerange{\pageref{firstpage}--\pageref{lastpage}} \pubyear{2015}
\label{firstpage}

\maketitle

\begin{abstract}
We present an end-to-end, two-phase model for the origin of globular clusters (GCs). In the model, populations of stellar clusters form in the high-pressure discs of high-redshift ($z>2$) galaxies (a rapid-disruption phase due to tidal perturbations from the dense interstellar medium), after which the galaxy mergers associated with hierarchical galaxy formation redistribute the surviving, massive clusters into the galaxy haloes, where they remain until the present day (a slow-disruption phase due to tidal evaporation). The high galaxy merger rates of $z>2$ galaxies allow these clusters to be `liberated' into the galaxy haloes before they are disrupted within the high-density discs. This physically-motivated toy model is the first to include the rapid-disruption phase, which is shown to be essential for simultaneously reproducing the wide variety of properties of observed GC systems, such as their universal characteristic mass-scale, the dependence of the specific frequency on metallicity and galaxy mass, the GC system mass--halo mass relation, the constant number of GCs per unit supermassive black hole mass, and the colour bimodality of GC systems. The model predicts that most of these observables were already in place at $z=1$--$2$, although under rare circumstances GCs may still form in present-day galaxies. In addition, the model provides important constraints on models for multiple stellar populations in GCs by putting limits on initial GC masses and the amount of pristine gas accretion. The paper is concluded with a discussion of these and several other predictions and implications, as well as the main open questions in the field.
\end{abstract}


\section{Introduction} \label{sec:intro}
All massive galaxies ($M_{\rm star}>10^9~\msun$) in the local Universe host populations of globular clusters (GCs), gravitationally bound stellar systems that are typically old ($\tau\sim10^{10}~\yr$) and massive ($M\sim10^5~\msun$). The origin of GCs is a major unsolved problem on the interface between star and galaxy formation. In part, our understanding of GC formation is limited because most GCs in the Universe must have formed at redshifts $z>2$. At such distances, the sizes of star-forming giant molecular clouds (GMCs) are generally unresolved, implying that the physical processes leading to GC formation must be inferred indirectly. Important, independent examples of such constraints on the GC formation process exist from the formation of massive clusters in the local Universe, the observed conditions for star and cluster formation in high-redshift galaxies, and the present-day properties of GC populations.

In this work, we present a new model for GC formation that combines the current observational and theoretical constraints. This model is the first to provide an end-to-end (yet simple) description for the origin of GCs, from their formation at high redshift, through the relevant evolutionary processes, until the present day. It explains the observed properties of GC populations as the natural outcome of regular star and cluster formation in the high-redshift Universe.

Before discussing the origin of GCs further, we should define the term `globular cluster'. In the literature, a large variety of definitions has been used, based on metallicity (`metal-poor'), mass ($M=10^4$--$10^6~\msun$), age ($\tau\sim10^{10}~\yr$), location (`in the halo'), or chemical abundance patterns (`multiple stellar populations'). However, exceptions to all of these criteria can be given. For instance, GCs are known to exist with (super-)solar metallicities, with masses of $M<10^3~\msun$, with ages $\tau\sim5~\gyr$ \citep[or even as young as a few $100~\myr$,][]{schweizer98}, with kinematics and positions consistent with the host galaxy's bulge component, or without multiple stellar populations \citep[e.g.][]{harris96,dinescu99,carraro06,forbes10,walker11b}. In this work, we will therefore use a definition which will be shown to be physically more fundamental. The term `globular cluster' will be used to refer to any object satisfying the following condition.

{\it A gravitationally-bound, stellar cluster that in terms of its position and velocity vectors does not coincide with the presently star-forming component of its host galaxy.}

In this definition, the term `stellar cluster' excludes dark matter-dominated systems such as dwarf galaxies, and `the presently star-forming component' refers to the part of position--velocity space occupied by a galaxy's star-forming gas. Of course, a GC may pass through this component -- in this case, it coincides in position, but not in velocity space. It may also co-rotate with the star-forming component, but in a position above or below the galactic plane (as holds for part of the GC population of the Milky Way, see \citealt{frenk80}). We have deliberately omitted limits on the GC metallicity, mass, or age. However, it is one of the goals of this paper to show that the ranges of these quantities observed in GCs are implied by a combination of the above definition and the physics of cluster formation and evolution.

Early theories aiming to explain the origin of GCs often invoked the special conditions in the early Universe \citep[e.g.~high Jeans masses prior to the formation of the first galaxies or in young galaxy haloes, see][]{peebles68,fall85}. More recent models have begun to focus on the dynamical evolution of GCs within their host galaxy haloes in an attempt to reconstruct their present-day properties from an assumed initial population of stellar clusters \citep[e.g.][]{fall01,prieto08,tonini13,li14}. These theories generally omit a description of the actual GC formation process itself and may therefore be missing important physics. Another family of models aims to constrain GC formation physics using the chemical abundance variations observed in present-day GCs \citep[e.g.][]{decressin07,dercole08,conroy11,krause13b,bastian13}. These models often require new \citep[and thus far unobserved, see][]{bastian13b,kruijssen14c} physics to explain the observed abundance variations.

In most of the above cases, the formation of GCs is reverse-engineered using their present-day properties and/or their recent dynamical evolution. However, further constraints can be added by considering the formation environment of GCs. The old ages of most GCs imply that the majority of them must have formed at $z>2$, close to the peak of the cosmic star formation history at $z=2$--$3$ \citep[e.g.][]{hopkins06b}. Indeed, observations of galaxies at these redshifts reveal vigorously star-forming systems, with star formation rates (SFRs) exceeding several $10^2~\msun~\yr^{-1}$ \citep[e.g.][]{forsterschreiber09,daddi10b}. The gas content of these high-redshift galaxies has been surveyed in great detail over the past couple of years \citep[e.g.][]{genzel10,tacconi10,tacconi13,swinbank11}, revealing high gas densities ($\Sigma=10^2$--$10^4~\msun~\pc^{-2}$) and turbulent velocities ($\sigma=10$--$100~\kms$), implying gas pressures ($P/k\sim10^7~{\rm K}~\cmc$) some three orders of magnitude higher than in the discs of nearby galaxies such as the Milky Way \citep[e.g.][]{kruijssen13c}. This typical pressure falls nicely in the range required for GC formation \citep[$P/k=10^6$--$10^8~{\rm K}~\cmc$, see][]{elmegreen97}.

While the high-pressure conditions of star formation in $z>2$ galaxies may seem unique, analogous environments exist in the local Universe. Galaxy mergers, starburst dwarf galaxies, and galactic nuclei are all known to reach the extreme pressures seen regularly at high redshift \citep[e.g.][]{downes98,kruijssen13c}. Such high-pressure environments are almost universally seen to produce stellar clusters that are indistinguishable from GCs in terms of their masses and radii \citep{portegieszwart10,longmore14,kruijssen14c}. These `young massive clusters' (YMCs; with masses $M=10^4$--$10^8~\msun$, radii $r_{\rm h}=0.5$--$10~\pc$, and ages $\tau<1~\gyr$) had already been discovered in the 1980s \citep{schweizer82,schweizer87}, but their number increased spectacularly after the launch of the Hubble Space Telescope \citep[HST; see e.g.][]{holtzman92,schweizer96,whitmore99,bastian06}. These observations unequivocally showed that GC-like clusters are still forming in the local Universe wherever the conditions are similar to those seen in star-forming galaxies at the peak of the cosmic star formation history at $z=2$--$3$. The obvious difference between YMCs and GCs is that the former are generally still associated with the presently star-forming components of their host galaxies (thereby disqualifying them as GCs in the above definition), whereas the latter are not. In this paper, it is proposed that this separation occurs naturally during galaxy mergers and the accretion of satellite galaxies, both of which are common events in the context of hierarchical galaxy formation \citep{white78,white91}.

Given the ubiquity of GC-like cluster formation under high-pressure conditions, it is undesirable to devise `special' physical mechanisms to explain the origin of GCs. Instead, the primary question(s) to ask should be:
\begin{enumerate}
\item
Could the products of regular cluster formation in high-redshift galaxies have survived until the present day?
\item
If so, are these relics consistent with the properties of present-day GC populations?
\end{enumerate}
It is the aim of this work to address the above questions by constructing a simple model for (1) the formation of regular stellar clusters in the early Universe and (2) their survival until the present-day in the context of hierarchical galaxy formation. To do so, we combine the current observational and theoretical constraints on nearby YMC formation, high-redshift star formation, and present-day GC populations. While no fundamentally new physical mechanisms are added, this is the first time that an end-to-end model for the cosmological formation and evolution of massive clusters includes a physical description for the formation and early evolution of these clusters within the high-pressure environments of their host galaxies. We show that this initial phase plays an essential role in setting the properties of the surviving cluster population (its importance was also highlighted by \citealt{elmegreen10} and 
\citealt{kruijssen12c}). As will be discussed at length, the modelled stellar cluster populations at $z=0$ match a wide range of independent, observed properties of present-day GC populations, such as the GC mass function, specific frequency, and the constant GC population mass per unit dark matter halo mass.

Of course, this is not the first paper aiming to put GCs in the context of galaxy formation. However, previous efforts to explain the properties of GC populations have focused on:
\begin{enumerate}
\item
exploring special, high-redshift conditions in which GCs could conceivably have formed (e.g.~in protogalactic clouds with high Jeans masses, in galaxy mergers, or during reionisation, see \citealt{peebles68,fall85,ashman92,katz14});
\item modelling a Hubble time of dynamical evolution and evaporation of GCs in the haloes where they presently reside, with the aim of explaining the differences between observed young cluster populations and old GCs \citep[e.g.][]{prieto08,muratov10,katz14,li14};
\item following the GC population during hierarchical galaxy formation by assuming that some fraction of high-redshift star formation occurred in the form of GCs, without accounting for any cluster disruption \citep[e.g.][]{kravtsov05,tonini13}.
\end{enumerate}
The theory in this paper differs fundamentally from these previous approaches. Instead of exclusively focusing on special conditions for GC formation, disruption by tidal evaporation, or hierarchical galaxy growth, in this paper we take the ``Occam's Razor'' approach of combining all mechanisms that are known to affect the formation and evolution of stellar clusters, and to put these in the context of galaxy formation to verify if the properties of observed GC populations are retrieved. The resulting model is the first to include the environmental dependence of GC formation and early disruption by tidal shocks in the host galaxy disc. As mentioned above, we will show that this early phase is dominant in setting the properties of the $z=0$ GC population. In addition, we will improve on previous models for GC evaporation in galaxy haloes (which used the environmentally-independent expression of \citealt{spitzer87}) by following recent $N$-body simulations \citep[e.g.][]{baumgardt03,gieles08,lamers10} and accounting for the environmental dependence of GC evaporation due to the presence of a tidal field.

The paper is structured as follows. In \S\ref{sec:constr}, we will first summarise the main observational (\S\ref{sec:obs}) and theoretical (\S\ref{sec:phys}) constraints on GC formation. In \S\ref{sec:model}, we will combine these constraints in a simple, two-phase model for GC formation in which the violent conditions at high redshift promote GC formation due to (1) the high initial cluster masses that are attainable in high-pressure environments and (2) the rapid migration of (Y)MCs into the quiescent galaxy halo due to the high incidence of galaxy mergers at $z>2$, both of which enable the long-term survival of GCs. The predictions of the model are contrasted with a wide range of observational data, illustrating how the observed properties of the GC population have been sculpted by the violent conditions of their formation (\S\ref{sec:comp}). We also discuss the implications for our current understanding of GC formation and provide predictions for future work (\S\ref{sec:pred}). The main open questions in the field are discussed in \S\ref{sec:open}, and the summary and conclusions of this work are provided in \S\ref{sec:concl}.

\section{Current constraints on globular cluster formation} \label{sec:constr}
In this section, we summarise the constraints on GC formation \citep[also see the recent review of][]{kruijssen14c}. These constraints are drawn from the fields of nearby YMC formation, high-redshift star formation, and present-day GC populations, as well as the theory of cluster formation and evolution. Our new model will be described in \S\ref{sec:model} and is based on the constraints discussed in this summary. The comparison to observations in \S\ref{sec:comp} also focuses on the observables described here. The reader who is exclusively interested in the model is advised to skip this section and continue to \S\ref{sec:model} and \S\ref{sec:comp}.

\subsection{Observational constraints} \label{sec:obs}

\subsubsection{YMCs in the local Universe as analogues for GC formation} \label{sec:ymc}
Ever since the discovery of YMCs with masses similar to (or larger than) those of GCs in the 1980s and 1990s, numerous authors have suggested that these YMCs could be the progenitors of future GCs. Initially, the formation of these extreme stellar clusters was thought to be exclusive to galaxy mergers or merger remnants \citep[e.g.][]{ashman92}, but YMCs have now been seen in a variety of high-pressure environments, including starburst dwarf galaxies and galaxy centres \citep[e.g.][]{figer99,figer02,anders04}. If GCs indeed formed as regular YMCs, then the characterisation of the YMC formation process in the local Universe provides insight into the formation of GCs.

There is a wide range of literature on YMC formation \citep[see e.g.][]{portegieszwart10,longmore14}. The observational constraints most relevant in the context of GC formation are summarised here.
\begin{enumerate}
\item
YMCs are seen to form through the hierarchical merging of smaller structures \citep{elmegreen01,bastian07,sabbi12,gouliermis14,rathborne15,walker15} on a short ($\sim\myr$) time-scale \citep{elmegreen00,bastian13b,cabreraziri14}. After this time, no gas is left within the YMC due to a combination of gas consumption in star formation and gas expulsion by stellar feedback and the YMC resides in virial equilibrium \citep{rochau10,cottaar12,henaultbrunet12,clarkson12}.
\item
Only some fraction of all star formation occurs in bound stellar clusters \citep{lada03}, which is referred to as the the cluster formation efficiency (CFE or $\Gamma$, \citealt{bastian08}). Within the hierarchical density structure of the interstellar medium (ISM), the highest-density peaks achieve the highest star formation efficiencies (SFEs or $\epsilon$), allowing them to remain gravitationally bound when star formation is halted by feedback \citep{elmegreen97,elmegreen08,kruijssen12d,wright14}. Because the mid-plane gas density increases with the gas pressure, the CFE is seen to increase with the pressure too, from $\Gamma\sim0.01$ at $P/k<10^4~\kcmc$ to $\Gamma\sim0.5$ at $P/k>10^7~\kcmc$ \citep{larsen00,goddard10,cook12,silvavilla13,adamo15b}.
\item
The initial cluster mass function (ICMF) follows a power law (${\rm d}N/{\rm d}M_{\rm i}\propto M_{\rm i}^\alpha$) with index $\alpha=-2$ for cluster masses $M>10^2~\msun$ \citep{zhang99,lada03,larsen09}. At the high-mass end, the ICMF has a truncation that is well-fit by an exponential function [${\rm d}N/{\rm d}M_{\rm i}\propto M_{\rm i}^\alpha\exp{(-M_{\rm i}/M_{\rm c})}$, where $M_{\rm c}$ is the truncation mass, see e.g.~\citealt{gieles06b,larsen09,bastian12}], which was originally proposed by \citet{schechter76} to describe the galaxy mass function.
\item
The ICMF truncation mass is proportional to the \citet{toomre64} mass $M_{\rm T}=\sigma^4/G^2\Sigma$ of the host galaxy disc \citep{kruijssen14c,adamo15b}, where $\sigma$ is the gas velocity dispersion and $\Sigma$ is the gas surface density. This mass-scale represents the two-dimensional Jeans mass, i.e.~the maximum mass-scale below which self-gravity can overcome galactic shear \citep{elmegreen83,kim01}, and determines the mass of the most massive GMC in a galaxy. The mass of the most massive cluster $M_{\rm T,cl}$ is subsequently obtained by multiplying the Toomre mass by the SFE and the CFE, i.e.~$M_{\rm T,cl}=\epsilon\Gamma M_{\rm T}$. The Toomre mass steeply increases with the pressure, hence the mass of the most massive cluster in a galaxy is a good proxy for the pressure in its formation environment. Indeed, nearby galaxies that produce clusters with masses similar to the most massive GCs ($M>10^6~\msun$) have ISM pressures similar to those seen at high redshift \citep{kruijssen14c}.
\item
YMCs have radii ($r_{\rm h}=0.5$--$10~\pc$) that are roughly independent of their masses \citep{larsen04b}. The variation of the initial cluster radius with the galactic environment is an open problem -- a slight decrease of the radius with the ambient density may exist. This would be expected if clusters born in strong tidal fields (or high-pressure environments) are more compact than those in weak tidal fields \citep[e.g.][]{elmegreen08}. Scaling the typical YMC radius in the Galactic disc to the strong tidal field of the Galactic centre does reproduce the small ($0.5~\pc$) radius of the Arches cluster, which resides within the central $\sim100~\pc$ of the Milky Way. While this comparison is certainly suggestive, this is an area where current constraints must be improved.
\end{enumerate}
The above points show that the formation of GC-like YMCs ($M\geq10^5~\msun$, see \S\ref{sec:comp} and \S\ref{sec:pred}) occurs naturally under high-pressure conditions, due to the elevated CFEs and Toomre masses in such environments.

\subsubsection{Star formation in high-redshift galaxies as the environmental conditions for GC formation} \label{sec:hiz}
Most GCs have ages in the range $\tau=10$--$13~\gyr$ \citep[e.g.][]{forbes10,vandenberg13}, indicating that the peak of cosmic GC formation must occur at redshift $z=2$--$6$. In this paper, a fiducial GC formation redshift of $z\sim3$ is adopted.\footnote{We emphasise that the formation redshift of $z\sim3$ is not rigid and we certainly do not mean to suggest that GC formation is restricted to $z=2$--$6$. It is only intended to represent a reasonable choice for the majority of the GC population. In fact, one of the main conclusions of this work will be that GCs are still forming in the present-day Universe.} Galaxies at these redshifts are too distant to observe the GC formation process in detail, but the global conditions for star formation in $z>2$ galaxies (i.e.~near the peak of the cosmic star formation history at $z=2$--$3$, \citealt{hopkins06b}) can provide further constraints on the formation of GCs \citep[e.g.][]{shapiro10,elmegreen10,kruijssen12c}. For instance, they can help answer whether the properties of the ISM in these galaxies promote the formation of GC-like YMCs.

Like for YMC formation, there exists a large body of literature on star formation in high-redshift galaxies. Here, we summarise the properties of $z>2$ galaxies relevant for GC formation.
\begin{enumerate}
\item
High-redshift galaxies with stellar masses $M_{\rm star}>4\times10^9~\msun$ are characterised by high gas fractions ($f_{\rm gas}\sim0.5$, \citealt{tacconi13}) and specific star formation rates (${\rm sSFR}\equiv{\rm SFR}/M_{\rm star}\sim3~\gyr^{-1}$, \citealt{bouche10,rodighiero11}), indicating that a large fraction of their stellar mass is in the process of being formed. The star formation rate per unit gas mass (or alternatively the gas depletion time $t_{\rm depl}\equiv{\rm SFR}/M_{\rm gas}$) is similar to that in nearby galaxies \citep[$t_{\rm depl}\sim10^9~\yr$,][]{bigiel11,schruba11,tacconi13,fisher14}.
\item
The metallicities of star-forming galaxies at $z\sim3$ are lower than those in the local Universe by a factor of $\sim5$, but the overall shape of the galaxy mass--metallicity relation is similar. In \citet{kruijssen14c}, we parameterised the mass--metallicity relations of \citet{erb06} and \citet{mannucci09} as
\bea
\label{eq:fehmstar}
\feh& \sim &-0.59 + 0.24 \log{\left(\frac{M_{\rm star}}{10^{10}~\msun}\right)} \\
\nonumber  &&- 8.03\times10^2 \left[\log{\left(\frac{M_{\rm star}}{10^{10}~\msun}\right)}\right]^2- 0.2(z-2) ,
\eea
for the redshift range $z=2$--$4$. This functional form suggests that at $z\sim3$, clusters with typical GC metallicities (i.e.~between $\feh\sim-1.6$ and $\feh\sim-0.5$, \citealt{peng06}) can form in galaxies with stellar masses $M_{\rm star}=10^8$--$10^{11}~\msun$.
\item
In high-redshift galaxies, the formation of stars and stellar clusters proceeds at high densities and pressures, with gas surface densities $\Sigma=10^2$--$10^{3.5}~\msun~\pc^{-2}$, SFR surface densities $\Sigma_{\rm SFR}=10^{-1}$--$10^{0.5}~\msun~\yr^{-1}~\kpc^{-2}$, and gas velocity dispersions $\sigma=10$--$100~\kms$ \citep{genzel10,swinbank12,tacconi13}, implying pressures of $P/k>10^7~\kcmc$. These conditions give rise to high CFEs ($\Gamma\sim0.5$, \citealt{kruijssen12d}) and Toomre masses ($M_{\rm T}\ga10^8~\msun$ and therefore $M_{\rm T,cl}\ga10^{6.5}$, \citealt{kruijssen14c}), both of which promote the formation of GC-like YMCs.
\item
In particular, the gas-rich discs of $z>2$ galaxies exhibit clumpy morphologies \citep{elmegreen05,genzel11,tacconi13}, suggesting widespread gravitational instabilities. These clumps have size-scales of $\sim1~\kpc$ and masses $M\ga10^8~\msun$, and may provide a natural formation environment for massive stellar clusters \citep{shapiro10}. Note that most gravitational instabilities form at large galactocentric radii \citep[out to several $\kpc$, e.g.][]{swinbank12}, implying that even the most massive clusters forming within them would require $>10^{10}~\yr$ to spiral in to the galaxy centre by dynamical friction.
\item
While the high-pressure conditions in gas-rich galaxies promote the formation of massive clusters, the correspondingly-high gas densities should also limit the long-term survival of stellar clusters due to the frequent and strong tidal perturbations by dense gas pockets. These `tidal shocks' have been shown to be the dominant cluster disruption agent in gas-rich environments, both in empirical studies \citep{lamers05} and theoretical work \citep[e.g.][]{lamers06a,elmegreen10b,kruijssen11}. Extrapolating the empirically-estimated lifetimes of clusters in local-Universe, high-pressure conditions \citep[e.g.][]{gieles05} suggest that $M\la10^6~\msun$ clusters would survive for $t_{\rm dis}\la10^9~\yr$ in $z>2$ galaxies.
\end{enumerate}
We see that even though massive cluster formation should take place under the high-pressure conditions of $z>2$ galaxies, the lifetimes of stellar clusters in such environments are very limited. In order to survive for a Hubble time, the GC progenitor clusters must therefore have escaped the gas-rich bodies of their host galaxies before they got disrupted (see \S\ref{sec:phys} below).

\subsubsection{Present-day GC populations as the outcome of the process we are seeking to understand} \label{sec:gc}
The product of the GC formation process and a Hubble time of GC evolution is readily observable in nearby galaxies. Any theory of GC formation should yield observables at $z=0$ consistent with the properties of present-day GC populations. There is a wealth of reviews discussing GCs in the nearby Universe \citep[e.g.][]{harris91,brodie06,gratton12}. The observables most relevant to this work are summarised below, again drawing from the most recent review of \citet{kruijssen14c}.
\begin{enumerate}
\item
The globular cluster mass function (GCMF) is peaked, with a near-universal peak mass of $M_{\rm peak}\sim2\times10^5~\msun$ \citep{jordan07}, in strong contrast with the Schechter-type ICMF seen for young stellar cluster populations (see \S\ref{sec:ymc}). Only in galaxies with present-day stellar masses $M_{\rm star}<10^{10}~\msun$, a slight decrease of the peak mass with decreasing galaxy mass is observed \citep{jordan07}. This weak trend is currently unexplained.
\item
The near-universal peak mass is often interpreted as the result of GC evaporation, which leads to the gradual disruption of low-mass GCs over a Hubble time \citep{vesperini01,fall01}, thereby turning the Schechter-type ICMF into the peaked GCMF. However, in this scenario the peak mass should vary both with the host galaxy and the location within the halo, because the evaporation rate increases linearly with the orbital angular velocity (see \S\ref{sec:phys} and e.g.~\citealt{vesperini97b,portegieszwart98,baumgardt03}). Such a strong environmental variation is is not observed \citep{vesperini03,jordan07,mclaughlin08}. This calls into question the importance of evaporation in setting the peaked shape of the present-day GCMF. Clearly, an end-to-end understanding of the emergence of the GCMF from the ICMF cannot be achieved without more detailed knowledge of the mass loss histories of GCs.
\item
The GCMF has an exponential truncation at the high-mass end \citep{fall01,kruijssen09b}, which depends strongly on the host galaxy mass. In Milky Way-mass galaxies the typical truncation mass is $M_{\rm c}\sim3\times10^6~\msun$, but it is an order of magnitude lower in $M_{\rm star}\sim10^9~\msun$ galaxies \citep{jordan07}. Considering that the truncation mass of the ICMF of young stellar cluster populations is proportional to the Toomre mass (see \S\ref{sec:ymc}, \citealt{kruijssen14c} and \citealt{adamo15b}), it is plausible that this variation reflects a trend of increasing Toomre mass with galaxy mass (and metallicity, see \S\ref{sec:hiz}) at the time of GC formation.
\item
The importance of GC evaporation for the present-day statistics of GC populations is called into question further by considering the number of GCs per unit stellar mass in the host galaxy [i.e.~the specific frequency $T_N\equiv N_{\rm GC}/(10^9~\msun)$] as a function of the galactocentric radius. It is shown in Figure~\ref{fig:tnfeh_obs} that the specific frequency strongly decreases with increasing metallicity (and hence the host galaxy mass at formation, see \S\ref{sec:hiz}). This may result in extemely high specific frequencies in metal-poor dwarf galaxies like the Fornax dSph, IKN, and WLM, where up to 20\% of low-metallicity ($\feh<-2$) stars can reside in GCs \citep{larsen12,larsen14}. However, at fixed metallicity, there is no dependence of the specific frequency on the present-day galactocentric radius \citep{harris02,lamers15}. This suggests a minor impact of evaporation (which would have required a clear radial variation of $T_N$, see \S\ref{sec:phys}) and challenges models aiming to explain the present-day properties of the GC population as the result of evaporation \citep[e.g.][]{fall01,li14}. Instead, it seems more likely that the specific frequency was set at the time of GC formation or during the early evolution of GCs, before they reached their present spatial configuration within galaxy haloes. In that case, the relation between $T_N$ and $\feh$ (or the host galaxy mass at formation) may be fundamental in setting other observed correlations between the properties of GC systems and their host galaxies (see below).
\begin{figure}
\center\includegraphics[width=\hsize]{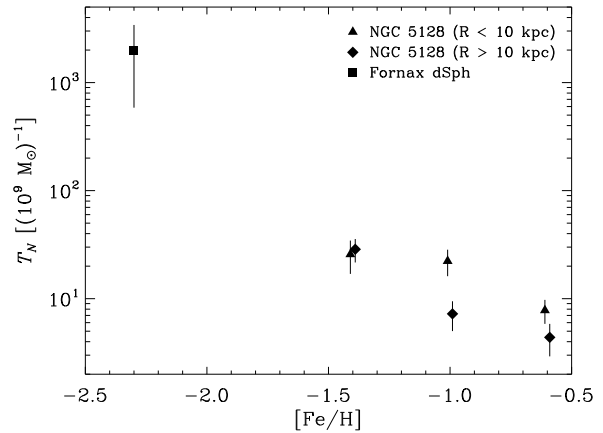}
\caption{This figure shows that the specific frequency of GCs does not depend on the present galactocentric radius and therefore must have been set during their formation or early evolution. Shown is the specific frequency $T_N$ as a function of metallicity $\feh$ for GCs in the inner region of NGC~5128 \citep[triangles, $R\la10~\kpc$; data from][]{harris02}, the outer region of NGC~5128 \citep[diamonds, $R\ga10~\kpc$; data from][]{harris02}, and in the Fornax dSph \citep[square; data from][]{larsen12}. The division into radial bins for the GCs in NGC~5128 shows that $T_N$ chiefly depends on $\feh$, exhibiting little variation with the radius at fixed $\feh$. The data point for the Fornax GC system is included solely to illustrate the metallicity trend to lower metallicities.} \label{fig:tnfeh_obs}
\end{figure}
\item
The total mass of a galaxy's GC population is a near-constant fraction of the galaxy's dark matter halo mass, i.e.~$\log{\eta}\equiv \log{(M_{\rm GC,tot}/M_{\rm h})}=-4.5$ in the halo mass range $M_{\rm h}=10^9$--$10^{15}~\msun$ \citep[e.g.][]{spitler09,georgiev10,harris13,hudson14,durrell14}. It is unclear if this linear relation between GC system mass and halo mass reflects a fundamental connection between dark matter and GCs. Even if such a relation initially existed, it must have been affected by a Hubble time of GC disruption.
\item
The GC colour and metallicity distributions are bimodal \citep[e.g.][]{zinn85,peng06}. While the spatial distribution and kinematics of the metal-rich GC sub-population bear some imprint of the host galaxy's spheroid, the metal-poor GCs are mainly associated with the stellar halo. Recent theoretical work has shown that the metallicity bimodality of GCs may emerge naturally from hierarchical galaxy formation \citep{tonini13}. In this scenario, metal-rich GCs are formed together with the main spheroid of the galaxy and metal-poor GCs accreted by the tidal stripping of satellite (dwarf) galaxies. However, this model uses the $z=0$ relation between galaxy mass and specific frequency (which is intimately related to the relation of Figure~\ref{fig:tnfeh_obs}) to initialise the GC population at the time of GC formation. It therefore omits the physics of GC formation and disruption.
\end{enumerate}

\subsubsection{Chemical abundance patterns} \label{sec:chem}
Next to the characteristics of the GC population, additional constraints on GC formation may potentially be obtained from their composition. GCs are observed to host multiple stellar populations in terms of their light element abundances \citep[e.g.][]{gratton12}, such as an anti-correlation between the Na and O abundances of their constituent stars. This anti-correlation is almost a defining feature of GCs, as it has been found in all considered GCs \citep[with masses $M\geq3\times10^4~\msun$, e.g.][]{gratton12}. These abundance patterns have recently been heavily exploited to construct scenarios for the formation of GCs, almost exclusively from a stellar evolutionary perspective.

Three main models aiming to explain the observed abundance patterns exist (see \citealt{kruijssen14c} for a detailed discussion), but each still have substantial problems to overcome -- no definitive model for the multiple stellar populations in GCs has been achieved \citep{bastian15}. The viability of the existing models should be tested by comparing their predictions and assumptions to independent constraints on GC formation, i.e.~by not only considering stellar-evolutionary constraints, but also including what is known about GC formation and evolution. Here, we summarise the chief models for later reference in \S\ref{sec:pred}.
\begin{enumerate}
\item
Enrichment by winds from asymptotic giant branch (AGB) stars \citep[e.g.][]{dercole08,conroy11}. This model reproduces the abundance variations with the formation of a second generation of stars after enrichment by AGB ejecta ($>30~\myr$ after the formation of the first generation).
\item
Enrichment by winds from fast-rotating massive stars \citep[FRMSs; e.g.][]{decressin07,krause13b}. This model reproduces the abundance variations with the formation of a second generation of stars from retained FRMS ejecta ($<10~\myr$ after the formation of the first generation).
\item
Enrichment by the early disc accretion (EDA) of winds from FRMSs and massive interacting binaries \citep[MIBs;][]{demink09,bastian13}. This model reproduces the abundance variations by the sweeping up of enriched ejecta from FRMSs and MIBs with the protoplanetary discs around pre-existing low-mass stars. Out of the three models discussed here, it is the only model that does not require the formation of a second generation of stars.
\end{enumerate}
The AGB and FRMS models require GCs to have been 10-100 times more massive at birth to create an enriched gas reservoir sufficiently massive reproduce to the observed, 1:1 ratio between enriched and unenriched stars (the {\it mass budget problem}). In addition, the AGB model requires substantial pristine gas accretion to turn the abundance patterns resulting from stellar evolution into the observed ones (the {\it gas accretion problem}). By contrast, the FRMS model needs to retain gas for $\sim30~\myr$ after the formation of the first generation despite vigorous feedback from massive stars (the {\it gas retention problem}). Finally, the EDA model requires protoplanetary discs to survive for 5--10~$\myr$ (the {\it disc lifetime problem}), which may be longer than discs survive in nearby star-forming regions \citep{haisch01,bell13}, especially at the extreme stellar densities of GCs \citep{adams04,dejuanovelar12,rosotti14}.

We will revisit these problems in \S\ref{sec:pred}, by addressing questions such as how much more massive GCs could have been at birth given their disruption histories, or how much mass young GCs could have accreted under the conditions of high-redshift galaxies.

\subsection{Physics of globular cluster formation and evolution} \label{sec:phys}
We now briefly summarise the physical processes governing the formation of GCs and their evolution until the present day.
\begin{enumerate}
\item
Whereas this work has the specific aim of testing whether GCs may be the descendants of regular YMC formation at high redshift, other GC formation theories exist too. It was proposed that GC from in mergers \citep{ashman92}, but it is now clear that the universally-high Toomre mass in the $z>2$ Universe enables the formation of GC-like YMCs in normal disc galaxies too \citep[e.g.][]{shapiro10}. Likewise, recent work \citep[e.g.][]{forbes10,georgiev10,conroy11b} seems to rule out that GC formation is the result of a high Jeans mass following recombination \citep{peebles68}, thermal instabilities in hot galaxy haloes \citep{fall85}, or star formation before reionization within individual dark matter haloes \citep[e.g.][]{peebles84,bekki06}. However, a small fraction of the GC population may represent the former nuclei of tidally stripped dwarf galaxies \citep{mackey05,lee09,hartmann11}. The fraction of GCs formed this way is likely small, i.e.~less than 15\% \citep{kruijssen12b}.
\item
Numerical simulations of YMC formation have advanced greatly in the last decade (see \citealt{kruijssen13} for a recent review). These models now reproduce the observational picture of the hierarchical fragmentation of the ISM \citep[e.g.][]{tilley07,bonnell08,krumholz12c} and the merging of stellar aggregates \citep{maschberger10,fujii12}, in which gravitationally bound structure naturally arises at the highest-density peaks \citep[e.g.][]{offner09,kruijssen12,girichidis12b}. On a galactic scale, this model of cluster formation in high-density peaks within a hierarchical ISM naturally reproduces the observation that the fraction of star formation occurring in bound clusters (the CFE) increases with the gas pressure or surface density \citep{kruijssen12d}. Likewise, the emergence of the power-law ICMF emerges naturally from the fractal structure of the ISM \citep{elmegreen96}, with a high-mass truncation that is set by the Toomre mass of the host galaxy and increases with the pressure \citep{kim01,kruijssen14c,adamo15b}. It was shown by \citet{krumholz05} that the Toomre mass can be expressed in terms of simple observables, i.e.~the gas surface density, angular velocity, and Toomre $Q$ disc stability parameter, if the assumption is made that star formation in galaxies occurs in discs residing in hydrostatic equilibrium.
\item
While high gas pressures and densities may promote the formation of bound stellar clusters by increasing the CFE and the ICMF truncation mass, they also effectively destroy bound structure by tidal perturbations (often referred to as `impulsive' or `tidal' shocks) due to encounters with GMCs \citep[e.g.][]{spitzer58,spitzer87,ostriker72,gieles06}. This concept of a high disruption rate at young cluster ages (named the `cruel cradle effect' by \citealt{kruijssen11}; also see \citealt{elmegreen10b}) is of key importance during the early phase of GC evolution within the host galaxy disc, particularly in high-pressure environments \citep[e.g.][]{elmegreen10,kruijssen14b}. Under such conditions, other disruption mechanisms such as evaporation or dynamical friction destroy stellar clusters more slowly than tidal shocks by at least an order of magnitude \citep{kruijssen11,kruijssen12c}.
\item
The rapid disruption of stellar clusters by tidal shocks implies that YMCs are unlikely to survive for cosmological time-scales unless they migrate out of the disc into the host galaxy halo. Present-day GCs are not associated with their host galaxies' gas reservoirs, illustrating that GCs must have gone through a migration event.\footnote{Our use of the term `migration' strictly refers to the escape of GCs from the part of (6D) position-velocity space that is occupied by the gas.} This could be caused by several mechanisms, but numerical simulations of the formation and/or disruption of clusters during hierarchical galaxy formation suggest that the most obvious redistribution of GCs takes place when galaxies merge or are tidally stripped by a more massive galaxy \citep{kravtsov05,prieto08,kruijssen12c,rieder13}.\footnote{A more speculative scenario could be that GC migration is facilitated by tidal heating or secular evolutionary processes such as dynamical interactions with the gas clumps seen in high-redshift galaxies.} These processes `liberate' GCs from the host galaxy disc into the halo. The migration time-scale is set by the rate of galaxy mergers with similar-mass galaxies (major mergers) or more massive galaxies (accretion events), which is $t_{\rm merge}=10^8$--$10^{10}~\yr$ at $z\sim3$ \citep[e.g.][]{genel09} and increases with cosmic time and galaxy mass \citep[e.g.][]{fakhouri10}. As a result, the migration rate of GCs into low-density environments per GC must have peaked at high redshift and at low galaxy masses.
\item
Current numerical simulations of cluster formation and disruption in a galactic context all find that the survival probabilities of pre-existing clusters are indeed greatly increased by the redistribution of matter during galaxy mergers and hierarchical galaxy growth \citep{kruijssen12c,renaud13,rieder13}. Only the clusters formed at the peak of a merger-induced starburst experience greatly enhanced levels of disruption due to the high gas densities \citep{kruijssen12c}. As such, the surviving clusters in the merger remnant predate the main starburst and typically have ages up to $200~\myr$ older than the field stars formed in a merger \citep{kruijssen11}. Note that mergers also play a key role in setting the properties of the composite GC populations observed at $z=0$, of which the widely different metallicities and binding energies suggest that they formed in a variety of galaxies that merged to form the present-day system \citep[e.g.][]{searle78,forbes01,muratov10,tonini13}.
\item
Once GCs reside in the host galaxy halo, they slowly dissolve by evaporation \citep{spitzer87,fall01} at a rate that depends on the strength of the tidal field and hence on the galactic environment (see \citealt{baumgardt03,gieles08}, as well as a more extensive discussion of this point in \S3.5 of \citealt{kruijssen14c}). For a flat rotation curve, the disruption time-scale due to tidal evaporation scales linearly with the galactocentric radius \citep{baumgardt03}. As highlighted in \S\ref{sec:gc}, evaporation therefore cannot explain the observed insensitivity of the GCMF peak mass and the specific frequency to the host galaxy mass or the galactocentric radius (at fixed metallicity), unless the strong environmental dependence of evaporation is ignored \citep[as in e.g.][]{fall01,prieto08,katz14,li14}. Secondary disruption mechanisms such as bulge or disc shocks \citep{gnedin99b} are environmentally dependent too. If most GCs are the products of regular YMC formation in high-redshift galaxies, both the shape of the GCMF and the specific frequency must therefore have been set at early times. As will be shown in \S\ref{sec:comp}, the best candidate is the rapid disruption phase within the host galaxy disc, i.e.~before the GCs attained their current spatial distribution within galaxy haloes. This conclusion is supported further by the fact that the present-day mass loss rates of GCs are too low to facilitate the evolution of a Schechter-type ICMF into the peaked present-day GCMF within $\sim10^{10}~\yr$ \citep{fall01,vesperini03,kruijssen09b}. While GC evolution in galaxy haloes cannot have strongly affected the present-day masses and numbers of GCs, it does play a key role in setting the radii of GCs \citep{gieles11b,madrid12}. Recent observational work shows that this structural evolution of GCs within galaxy haloes is still ongoing \citep{baumgardt10}.
\begin{figure*}
\center\includegraphics[width=\hsize]{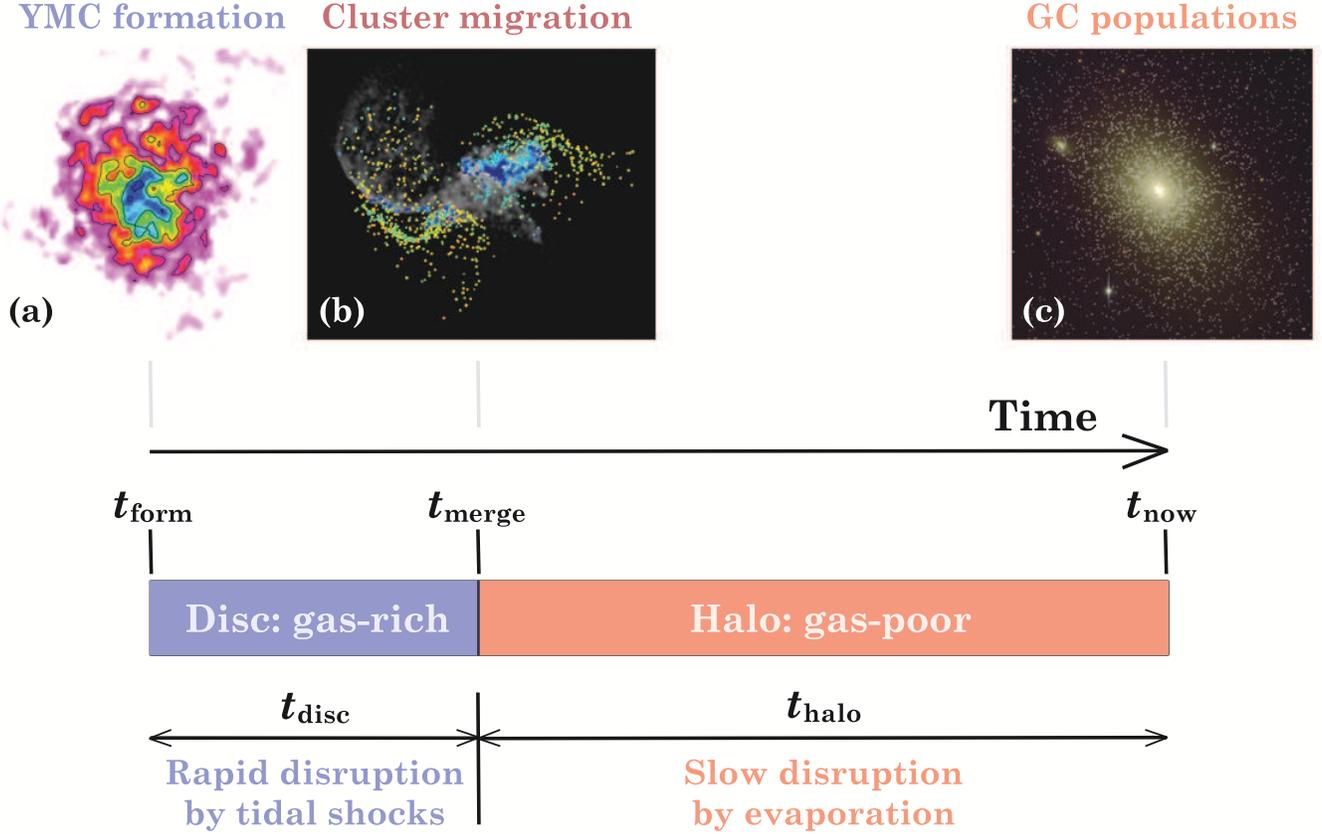}
\caption{This figure shows the time-line of the two-phase `{\it shaken, then stirred}' model for GC formation that is presented in \S\ref{sec:model}. At a time $t_{\rm form}$, stars and stellar clusters are formed through the regular star formation process in the gas-rich and high-pressure discs seen at high redshift (see panel~a). During their subsequent evolution within the host galaxy disc (Phase~1), the clusters undergo rapid disruption by tidal shocks for a total duration $t_{\rm disc}$, until at a time $t_{\rm merge}$ the host galaxy undergoes a merger and the clusters migrate into the galaxy halo (see panel~b), thereby increasing their long-term survival chances and thus becoming GCs. The GCs subsequently undergo quiescent dynamical evolution in the galaxy halo (Phase~2), which is characterised by slow disruption by evaporation (see panel~c). This phase lasts for a total duration $t_{\rm halo}$ until the present day $t_{\rm now}$. Panel~(a) shows the clumpy $z=4.05$ galaxy GN20 from \citet{hodge12}, panel~(b) shows the migration of stellar clusters in a galaxy merger model from \citet{kruijssen12c}, and panel~(c) shows the GC population of the early-type galaxy NGC~4365 from the SLUGGS survey \citep{brodie14}. Panels~(a) and~(c) reproduced by permission of the AAS.} \label{fig:tschem}
\end{figure*}
\end{enumerate}

\section{Shaken, then stirred: a two-phase model for globular cluster formation} \label{sec:model}
Thus far, models aiming to explain the properties of the GC populations seen in the local Universe have focused mainly on the dynamical evolution of GCs in the haloes where they presently reside \citep{fall77,chernoff90,gnedin97,vesperini01,fall01,mclaughlin08,lamers10,muratov10,li14}, thereby omitting the early phase of GC formation and evolution within the gas-rich discs of their host galaxies. In this section, we use the available observational and theoretical evidence discussed in \S\ref{sec:obs} and \S\ref{sec:phys} to present a simple two-phase model for GC formation that includes this early phase of cluster formation and evolution as well as the migration of GCs into the galaxy halo and their subsequent, quiescent evolution until the present day. 

The model is illustrated graphically in Figure~\ref{fig:tschem} and is explained fully below. A summary is provided in \S\ref{sec:modsum}, where we also highlight the differences with respect to previous models of GC formation and evolution in the context of galaxy formation. In \S\ref{sec:comp}, the model results are compared to observations of present-day GC populations, after which the implications of the model as well as its testable predictions are discussed in \S\ref{sec:pred}.

The GC formation model presented here connects the formation of YMCs in the local Universe (\S\ref{sec:ymc}), the properties of high-redshift, star forming galaxies (\S\ref{sec:hiz}), the hierarchical growth of galaxies (\S\ref{sec:phys}), and the physical mechanisms governing cluster evolution until the present day (\S\ref{sec:phys}). The model is constituted by the following steps.

\subsection{Initialisation of the host galaxy and its cluster population} \label{sec:init}
At the time of GC formation, the properties of the host galaxy and its cluster population are initialised according to the observational and theoretical constraints on local YMC formation and high-redshift star formation.
\begin{enumerate}
\item
The progenitor clusters of GCs formed at a redshift $z$, which may vary with the galaxy mass and metallicity. For illustration, it is here chosen to be $z=3$ independently of the galaxy mass, corresponding to a look-back time of $t=11.7~\gyr$ \citep[consistent with the ages of GCs, e.g.][]{wright06,forbes10,vandenberg13}. In the context of Figure~\ref{fig:tschem}, we thus set $t_{\rm form}=0~\gyr$ and $t_{\rm now}=11.7~\gyr$. This is obviously a simplification given the extended formation histories of GC populations, which is made to keep the model as transparent as possible. In future work, we will add the complexity of extended star and cluster formation histories (see \S\ref{sec:open}).
\item
In the model, the properties of the GC population are derived as a function of the host galaxy dark matter halo mass at GC formation $M_{\rm h}$. Given some value of $M_{\rm h}$, the host galaxy stellar mass $M_{\rm star}$ follows from the commonly-used abundance matching model of \citet{moster13}. Their abundance matching technique makes the assumption that $M_{\rm star}$ is a monotonically increasing function of $M_{\rm h}$ in order to connect the observed galaxy mass function to the halo mass function from the Millennium Simulation \citep{springel05d} for the redshift range $z=0$--$4$. Analytically, the relation is given by
\begin{equation}
\label{eq:ms}
M_{\rm star}=A M_{\rm h}\left[\left(\frac{M_{\rm h}}{M_1}\right)^{B}+\left(\frac{M_{\rm h}}{M_1}\right)^{C}\right]^{-1} .
\end{equation}
At the adopted $z=3$, the constants are $A=0.033$, $B=-0.76$, $C=0.85$, and $M_1=3\times10^{12}~\msun$ \citep{moster13}. Given some stellar mass, the metallicity $\feh$ of the host galaxy (and hence of its stellar cluster population) is obtained through equation~(\ref{eq:fehmstar}) in \S\ref{sec:hiz}. This direct relation to metallicity is important, because it shows that the metallicities of GCs trace the masses of the galaxies that they formed in.
\item
Stellar clusters of metallicity $\feh$ are generated according to a Schechter-type ICMF:
\begin{equation}
\label{eq:icmf}
\frac{{\rm d}N}{{\rm d}M_{\rm i}}\propto M_{\rm i}^\alpha\exp{(-M_{\rm i}/M_{\rm c})},
\end{equation}
with index $\alpha=-2$, a lower mass limit of $M_{\rm min}=10^2~\msun$ \citep[e.g.][]{lada03,kruijssen14c}, and a truncation mass $M_{\rm c}$, normalised to a total mass in clusters of $\Gamma M_{\rm star}$. This is equivalent to assuming a roughly instantaneous formation of the host galaxy's stellar and cluster populations.
\item
The metallicity of the cluster population and its host galaxy is translated to an ICMF truncation mass $M_{\rm c}$ by using the relation between the mean metallicity of the present-day GC population and the GCMF truncation mass $M_{\rm c}$ (i.e.~combining the $\feh$--$M_B$ relation from Figure~13 of \citealt{peng06} with the $M_{\rm c}$--$M_B$ relation from Figure~16 of \citealt{jordan07}). We thus obtain an approximate relation of 
\begin{equation}
\label{eq:mc}
\log{(M_{\rm c}/\msun)}\sim 6.5+0.7\feh .
\end{equation}
Both the $\feh$--$M_B$ and $M_{\rm c}$--$M_B$ relations are tightly defined, hence equation~(\ref{eq:mc}) should also have little scatter ($<0.3$~dex). The dependence on galaxy mass (and hence $\feh$) of the truncation mass must be physical in nature rather than a trivial statistical result of lower-mass galaxies producing fewer clusters. When excluding nuclear clusters, no dwarf galaxy ($M_{\rm star}<5\times10^8~\msun$) is known to host a GC more massive than $10^6~\msun$ \citep{georgiev10,kruijssen12b}, whereas the much more massive Milky Way has several such massive GCs without any of the Fe-spreads that would indicate a nuclear origin. This shows that even when adding up the GC populations of a large number of dwarf galaxies, no GCs as massive as in the Milky Way are found, which thereby substantiates the $M_{\rm c}$--$M_B$ relation of \citet{jordan07}.
\item
To obtain the gas surface density and the CFE $\Gamma$ at the time of GC formation, we use the relation between the truncation mass $M_{\rm c}$ and the Toomre mass $M_{\rm T}$, i.e.~by writing (also see \S\ref{sec:ymc}):
\begin{equation}
\label{eq:mc2}
M_{\rm c}\equiv M_{\rm cl,T}=\epsilon\Gamma M_{\rm T}=\epsilon\Gamma(\Sigma,Q,\Omega)\frac{\pi^4 G^2\Sigma^3 Q^4}{4\Omega^4},
\end{equation}
where $\Sigma$ is the gas surface density, $Q$ is the \citet{toomre64} stability parameter, and $\Omega$ is the angular velocity. The final equality in this expression assumes that young GCs formed in galaxy discs in hydrostatic equilibrium to express $M_{\rm T}$ in terms of $\Sigma$, $\Omega$, and $Q$ \citep[also compare to the expression given in \S\ref{sec:ymc}]{krumholz05}. We assume a Toomre stability parameter of $Q=1$, and use the SINS galaxy sample of $z=1.1$--$3.5$ galaxies \citep{forsterschreiber09} to obtain a power-law fit of $\Omega$ as a function of galaxy mass $M_{\rm star}$, yielding
\begin{equation}
\label{eq:omega}
\Omega = 0.1~\myr^{-1}\left(\frac{M_{\rm star}}{10^{10}~\msun}\right)^{0.073} ,
\end{equation}
for $M_{\rm star}=10^9$--$10^{11.5}~\msun$. At fixed $Q=1$, this relation implies that the mid-plane gas density increases with the host galaxy stellar mass as $\rho_{\rm ISM}\propto(\Omega/Q)^2\propto M_{\rm star}^{0.15}$ \citep[cf.][]{krumholz05}. Finally, we adopt a SFE of $\epsilon=0.05$ \citep[e.g.][]{lada03,evans09,kruijssen14c}, and use the CFE model of \citet{kruijssen12d} to obtain $\Gamma(\Sigma,\Omega,Q)$. For a given galaxy mass $M_{\rm star}$, we then numerically solve the implicit relation of equation~(\ref{eq:mc2}) to obtain $\Sigma$ (and hence $\Gamma$ and $M_{\rm T}$).
\item
A relation should be adopted between the initial cluster masses and their half-mass radii. For simplicity, we adopt a constant radius across the entire cluster population, which is consistent with nearby YMCs (see \S\ref{sec:ymc}), and assume that during the initial phase of cluster evolution in the host galaxy disc, the half-mass radius is proportional to the mean tidal radius of the cluster population, i.e.~$r_{\rm t}\propto\Omega^{-2/3}$. As shown by equation~(\ref{eq:omega}), the galaxies under consideration here have a typical angular velocity of $\Omega=0.1~\myr^{-1}$ as opposed to $\Omega=0.025~\myr^{-1}$ in the solar neighbourhood and other nearby spiral galaxies \citep[e.g.][]{kennicutt98b}. The characteristic radius of nearby YMCs of $r_{\rm h}=3.75~\pc$ \citep{larsen04b} thus becomes $r_{\rm h}=1.5~\pc$ in cluster-forming galaxies at $z\sim3$.
\end{enumerate}
In summary, we choose a halo mass $M_{\rm h}$ and GC formation redshift (taken to be $z=3$) to initialise a galaxy with a known stellar mass $M_{\rm star}$, metallicity $\feh$, angular velocity $\Omega$, gas surface density $\Sigma$, and mid-plane gas volume density $\rho_{\rm ISM}$. This galaxy hosts a young stellar cluster population with a known ICMF and CFE, and we make a physically-motivated assumption for the typical cluster radii $r_{\rm h}$.

\subsection{Shaken by the cruel cradle effect: cluster evolution during the rapid-disruption phase} \label{sec:shake}
After the formation of the cluster population, the clusters evolve within the host galaxy disc in which they formed (Phase~1). During this phase, clusters are disrupted by the tidal injection of energy during encounters with GMCs (i.e.~tidal shocks) in the gas-rich, high-pressure environment.

We describe the disruption process using the model of \citet[equation~39 -- also see \citealt{gieles06}]{kruijssen12d}, which for the assumptions and definitions made in the previous steps (e.g.~a gas disc in hydrostatic equilibrium) can be expressed in terms of a mass loss rate
\begin{equation}
\label{eq:dmdtcce}
\left(\frac{{\rm d}M}{{\rm d}t}\right)_{\rm cce}=-\frac{M}{t_{\rm cce}},
\end{equation}
on a disruption time-scale
\begin{equation}
\label{eq:tcce}
t_{\rm cce}=t_{\rm 5,cce}\left(\frac{f_\Sigma}{4}\right)^{-1}\left(\frac{\rho_{\rm ISM}}{\msun~\pc^{-3}}\right)^{-3/2}\left(\frac{M}{10^5~\msun}\right)\phi_{\rm ad}^{-1} ,
\end{equation}
where the subscript `cce' refers to the `cruel cradle effect' (see \S\ref{sec:phys}), $t_{\rm 5,cce}=176~\myr$ is the proportionality constant that very weakly depends on a cluster's structural parameters but is taken to be constant here, $f_\Sigma\equiv\Sigma_{\rm GMC}/\Sigma$ is the ratio between the GMC surface density and the mean gas surface density, and $\phi_{\rm ad}$ is the `adiabatic correction' \citep{weinberg94b}, which accounts for the absorption of the tidally injected energy by adiabatic expansion in sufficiently dense clusters (see below). Each quantity has been normalised to the typical numbers under consideration here.\footnote{Note that the linear dependence on cluster mass $M$ is actually one on the half-mass density $\rho_{\rm h}$, but a constant radius was assumed in \S\ref{sec:init} above.}

Equation~(\ref{eq:tcce}) shows that clusters are disrupted most rapidly in galaxies with high absolute ISM densities as well as high surface density contrasts between the GMCs and the diffuse ISM. Because the typical disruption time $t_{\rm 5,cce}$ is much shorter than a Hubble time, the long-term survival of clusters is only possible if they migrate out of the host galaxy disc on a short time-scale. In our model, cluster migration is facilitated by galaxy mergers (see below) and hence the formation of long-lived GCs requires a high merger rate, as is seen at high redshift. Specifically, the long-term survival of clusters with masses up to several $10^5~\msun$ typically requires that migration should take place within $\sim1~\gyr$, comparable to the typical merger time-scale at $z\sim3$. 

Determining $f_\Sigma$ in equation~(\ref{eq:tcce}) requires some knowledge about the structure of the ISM. By equating the turbulent pressure to the hydrostatic pressure in a self-gravitating GMC, \citet{krumholz05} derive an expression for the GMC density in an equilibrium disc, which in combination with the Toomre mass yields the GMC radius and hence its surface density $\Sigma_{\rm GMC}$. Adopting a typical GMC virial ratio $\alpha_{\rm vir}=1.3$ we obtain
\begin{equation}
\label{eq:fsigma}
f_\Sigma=3.92\left(\frac{10-8f_{\rm mol}}{2}\right)^{1/2} ,
\end{equation}
where the molecular gas fraction $f_{\rm mol}$ is a function of $\Sigma$ as defined in equation~(73) of \citet{krumholz05}, yielding almost an entirely molecular ISM ($f_{\rm mol}\sim1$) at the surface densities considered here (see below) and hence the GMC-to-diffuse gas surface density ratio is $f_\Sigma\sim4$.

Finally, the parameter $\phi_{\rm ad}$ in equation~(\ref{eq:tcce}) is the adiabatic correction, which accounts for the inefficient conversion of the tidally injected energy to dynamical mass loss for tidal shocks taking place on a time-scale $t_{\rm pert}$ longer than the cluster's crossing time $t_{\rm cr}$ -- in such a case, (part of) the energy is absorbed by the adiabatic expansion of the cluster. We use the definition of \citet{weinberg94b} for a time-scale ratio $\phi_t\equiv t_{\rm pert}/t_{\rm cr}$ and assume the definition of $\phi_t$ in terms of the cluster density $\rho_{\rm h}$ and the ambient gas density $\rho_{\rm ISM}$ from equation~(37) of \citet{kruijssen12d}:
\begin{equation}
\label{eq:phiad}
\phi_{\rm ad}=(1+\phi_t^2)^{-3/2}=\left[1+9\left(\frac{\rho_{\rm h}/\rho_{\rm ISM}}{10^4}\right)\right]^{-3/2} .
\end{equation}
For the adopted half-mass radius of $r_{\rm h}=1.5~\pc$ and at an ISM density of $\rho_{\rm ISM}\sim3~\msun~\pc^{-3}$, the adiabatic dampening of tidally injected energy becomes important (i.e.~$\phi_t>1$) for cluster masses $M>10^5~\msun$, which substantially increases their survival fractions and may prevent disruption altogether for $M>{\rm several}~10^5~\msun$.

For clusters in the initial mass range $M_{\rm i}=10^2$--$10^8~\msun$, we numerically integrate equation~(\ref{eq:dmdtcce}) to obtain cluster masses $M$ during the rapid-disruption phase at some time $t$ after they formed. The cluster mass function at time $t$ is obtained by converting the ICMF of equation~(\ref{eq:icmf}) through conservation of the number of clusters:
\begin{equation}
\label{eq:cmf}
\frac{{\rm d}N}{{\rm d}M}=\frac{{\rm d}N}{{\rm d}M_{\rm i}}\frac{{\rm d}M_{\rm i}}{{\rm d}M} .
\end{equation}

In summary, given a cluster mass spectrum and the host galaxy ISM properties from \S\ref{sec:init}, the cluster masses are evolved by accounting for tidal shocks. Other than a small set of parameters discussed in \citet{kruijssen12d} that only weakly affect the mass loss rate and depend on e.g.~the structural properties of the clusters, this step of the model has no free parameters. The main uncertainty is the assumption that cluster radii are independent of the cluster mass, which is discussed in more detail in \S\ref{sec:open}.

\subsection{Cluster migration into the galaxy halo by galaxy mergers} \label{sec:migr}
The rapid disruption phase continues until the clusters migrate into the galaxy halo. The migration agent may be any process that facilitates cluster migration, such as secular evolution within the thick, gas-rich host galaxy disc, tidal heating, rapidly changing gravitational potentials due to galaxy-wide feedback \citep{pontzen12,maxwell14}, or (major or minor) galaxy mergers. In this model, we assume that the migration of clusters into the halo results from galaxy mergers. It is shown in \S\ref{sec:comp} that at $z\sim3$, mergers indeed occur frequently enough to enable the survival of a population of young GCs.

The merger time-scale of the host galaxy is obtained using the galaxy halo merger rates from \citet[see their Figure~3 in particular]{genel09}, who have used the Millennium Simulation to derive the merger rate per unit progenitor halo, time, and galaxy mass ratio as a function of halo mass. We integrate the merger rate over all mass ratios $x\equiv M_{\rm host}/M_{\rm other}<3$ (where $x<1$ indicates that the host galaxy is merging with a more massive galaxy). This mass ratio range includes major mergers (defined as $1/3<x<3$), as well as the cannibalism of the host galaxy by a more massive galaxy (defined as $x<1/3$). Both types of event redistribute clusters into the halo -- in the first case ($1/3<x<3$) this is done by violent relaxation, whereas in the second case ($x<1/3$) the redistribution takes place by the tidal stripping of the clusters from the host galaxy before it coalesces with the more massive galaxy. Mergers with $x>3$ are excluded, because these represent cases where the GC host galaxy merges with a much less massive galaxy, which are unlikely to cause a morphological transformation strong enough for the clusters to migrate into the halo \citep[e.g.][]{hopkins09}.

Inverting the merger rate resulting from the mass ratio integration yields the merger time-scale, which reflects the cluster migration time-scale in the model and hence the represents the duration of the first, rapid disruption phase $t_{\rm disc}\equiv t_{\rm merge}-t_{\rm form}$ (cf.~Figure~\ref{fig:tschem}). \citet{genel09} provide merger rates at $z=3$ for galaxy halo masses $M_{\rm h}=\{10^{11.3},10^{12.5},10^{13.5}\}$, which are the halo masses for which the model results will be shown below (see Table~\ref{tab:init} in \S\ref{sec:example}).

The only free parameter in this step of the model is the maximum halo mass ratio required for cluster migration into the halo (taken to be $x<3$). As stated above, this choice is well-motivated, but we have also verified that the exact choice does not strongly affect the merger time-scales. This can be inferred already by visual inspection of Figure~3(b) of \citet{genel09}, in which mass ratios with $x>3$ represent the clear minority of mergers for halo masses $M_{\rm h}<10^{13}~\msun$. The time-scales themselves follow directly from the Millennium Simulation, and therefore do not represent free parameters in the context of this model.

\subsection{Stirred by evaporation: cluster evolution during the slow-disruption phase} \label{sec:stir}
After their migration into the galaxy halo, the clusters predominantly lose mass due to evaporation and thus undergo quiescent evolution until the present day (Phase~2).\footnote{The mass loss rate due to evaporation is at least five times higher than disc shocking \citep[e.g.][]{gnedin97,dinescu99,kruijssen09,kruijssen09b}. Bulge shocking can be more important and could be modelled by introducing a factor $1-e$ in equation~(\ref{eq:tevap}), where $e$ is the orbital eccentricity \citep{baumgardt03}. However, it is omitted in our model, because (1) there are poor constraints on the orbital eccentricity of GCs as a function of metallicity across the galaxy mass range of interest, and (2) for the median GC eccentricity in the Galactic halo ($e\sim0.5$, see \citealt{dinescu99}), the evaporative mass loss rates are only affected by a factor of $\sim2$. Given that we will show in \S\ref{sec:comp} that the evaporation phase is sub-dominant for setting the properties of the $z=0$ GC population, omitting this aspect is preferable over introducing a free parameter that hardly affects the results.} Following \citet{baumgardt01} and \citet{lamers05b}, their disruption in the galaxy halo can be written as a mass loss rate 
\begin{equation}
\label{eq:dmdtevap}
\left(\frac{{\rm d}M}{{\rm d}t}\right)_{\rm evap}=-\frac{M}{t_{\rm evap}},
\end{equation}
on a disruption time-scale
\begin{equation}
\label{eq:tevap}
t_{\rm evap}=t_{\rm 5,evap}\left(\frac{M}{10^5~\msun}\right)^\gamma ,
\end{equation}
where we adopt $\gamma=0.7$ \citep[e.g.][]{fukushige00,baumgardt01,lamers05,gieles08} and $t_{\rm 5,evap}\propto\Omega^{-1}$ is the characteristic disruption time-scale of a $M=10^5~\msun$ GC. The disruption time-scales of GCs in the Milky Way cover a broad range of $t_{\rm 5,evap}=1$--$100~\gyr$, depending on the orbital parameters \citep[e.g.][]{baumgardt03,kruijssen09}. This time-scale is $1$--$3$ orders of magnitude longer than during the rapid-disruption phase in the host galaxy disc (cf.~equation~\ref{eq:tcce}).

The characteristic time-scale for mass loss by evaporation $t_{\rm 5,evap}$ is selected such that the near-universal characteristic mass-scale of GCs is reproduced as a function of $\feh$ at $z=0$. As explained in \S\ref{sec:gcmf2} below, this requires $t_{\rm 5,evap}$ to decrease with the metallicity. For a flat rotation curve $t_{\rm 5,evap}\propto \Omega^{-1}\propto R$ and thus the equivalent assumption is that the GC metallicity $\feh$ decreases with $R$. As discussed in \S\ref{sec:gc}, GC populations indeed exhibit such a radial metallicity gradient, as metal-poor GCs reside at larger galactocentric radii $R$ than metal-rich ones.

The adopted disruption time-scales can be tested quantitatively by assuming a flat rotation curve and writing
\begin{equation}
\label{eq:t5}
t_{\rm 5,evap}=t_{5,\odot}(R/8.5~\kpc),
\end{equation}
with $t_{5,\odot}=33.8~\gyr$ for $\gamma=0.7$ \citep{kruijssen09}. Using the 2010 edition of the \citet{harris96} catalogue of Galactic GCs, a least-squares fit to the observed gradient of the GC metallicity with disruption time (and hence radius through equation~\ref{eq:t5}) is
\begin{equation}
\label{eq:fehobs}
\feh_{\rm obs}=-1.11-0.46\log{(t_{\rm 5,evap}/10~\gyr)}.
\end{equation}
In the model, the adopted values of $t_{\rm 5,evap}$ for $\feh=\{-1.1,-0.7,-0.6\}$ are $t_{\rm 5,evap}/\myr=\{15.8,2.5,1.3\}$ (see \S\ref{sec:example}), implying a least-squares gradient of:
\begin{equation}
\label{eq:fehmodel}
\feh_{\rm model}=-1.03-0.50\log{(t_{\rm 5,evap}/10~\gyr)}.
\end{equation}
The very similar slopes and normalisations of the observed and adopted $t_{\rm 5,evap}$--$\feh$ relations in equations~(\ref{eq:fehobs}) and~(\ref{eq:fehmodel}) show that our assumed disruption time-scales give an accurate representation of the Galactic GC population. Even though $t_{\rm 5,evap}$ was chosen to reproduce the observed characteristic GC mass at $z=0$, we therefore use a relation that is quantitatively consistent with the theoretically-predicted disruption time-scales given the observed metallicity gradient of Galactic GCs.

It is plausible that the $t_{\rm 5,evap}$--$\feh$ (or $R$--$\feh$) gradient can be interpreted as an imprint of the binding energy at formation, and hence of the galaxy-mass (and hence metallicity) dependent redistribution of GCs in mergers and/or tidal stripping at $t_{\rm merge}$. In that case, this relation between $\feh$ and $t_{\rm 5,evap}$ may hold universally in galaxies where the decay of GC orbits by dynamical friction is negligible (also see \S\ref{sec:gradient}).

Perhaps surprisingly, an interesting corollary of the above is that the large-scale tidal field currently experienced by GCs reflects that at earlier times, despite mixing during the further hierarchical assembly of galaxies. After their initial migration, GCs may be stripped from their host galaxy if the host galaxy becomes a satellite of a more massive halo. However, this does not affect their disruption time-scales, as the stripping takes place when the density enclosed by the satellite's orbit in the more massive halo ($\rho_{\rm tid,sat}\propto\Omega_{\rm sat}^2$) equals the density enclosed by the GC's orbit within the satellite ($\rho_{\rm tid,GC}\propto\Omega_{\rm GC}^2$). As a result, the angular velocity of the GC does not change when it is stripped from its host galaxy, nor does stripping alter its disruption time-scale $t_{\rm 5,evap}\propto\Omega^{-1}$. The only exceptions to this rule occur if the satellite is on a very eccentric orbit, which is then inherited by the GC, or if the galaxy merger mass ratio is $x\sim1$ (which \citealt{genel09} show is rare). The present-day disruption time-scales of GCs may thus be taken as a proxy for their disruption time-scales shortly after their initial migration into the halo (as was shown numerically by \citealt{rieder13}).

After having completed the numerical integration of equation~(\ref{eq:dmdtcce}) until the time of cluster migration $t_{\rm merge}$, we continue evolving the cluster population by integrating equation~(\ref{eq:dmdtevap}) to obtain cluster masses $M$ during the slow-disruption phase. The cluster mass function at that any time $t>t_{\rm merge}$ is again obtained by converting the ICMF of equation~(\ref{eq:icmf}) through conservation of the number of clusters as in equation~(\ref{eq:cmf}).

As indicated in \S\ref{sec:init}, this evolved cluster population in the host galaxy disc still has a single metallicity. However, it is straightforward to account for the hierarchical growth of galaxies and their GC systems by constructing a metallicity-composite population. This is done by carrying out a summation of these single-metallicity (i.e.~single-progenitor galaxy) models with weights given by any desired (e.g.~observed) distribution function of cluster metallicities.

In summary, given an evolved cluster mass spectrum at the end of the rapid-disruption phase from \S\ref{sec:shake}, the cluster masses are evolved by accounting for tidal evaporation. This step of the model is carried out in direct accordance with the results of detailed $N$-body simulations of dissolving clusters. The only free parameter (or rather choice) is the relation between metallicity and the characteristic evaporation time-scale. As explained above and in \S\ref{sec:gcmf2}, this relation has been chosen to reproduce the peak mass of the GC mass function at $z=0$, which we show above is consistent with the observed metallicity gradient of GCs in the Milky Way (and plausibly in other galaxies, see \S\ref{sec:gradient}).

\subsection{Summary of the model and comparison to previous work} \label{sec:modsum}
The simple model presented in this section provides an end-to-end description of GC formation and evolution in the context of galaxy formation. For a given halo mass, a young stellar cluster population forms within a normal, star-forming disc galaxy with known stellar mass, metallicity, ISM properties, and cluster population properties set by the halo mass, at a fiducial redshift $z=3$ (corresponding to the median formation redshift of Galactic GCs). The clusters then undergo an evolutionary phase characterised by rapid disruption by tidal perturbations from molecular clouds and clumps in the host galaxy disc. This phase continues until the host galaxy halo merges with another halo that has at least $1/3$ of the host galaxy's halo mass (i.e.~the mass ratio is $x\equiv M_{\rm host}/M_{\rm other}<3$), which in the model leads to the migration of the clusters into the halo of the merger remnant (here assumed to occur instantaneously, either by a major merger if $x\sim1$ or by tidal stripping if $x<1/3$). The (progenitor halo) merger time-scale is taken from the Millennium Simulation. The further evolution of the cluster population until the present day is characterised by slow disruption due to tidal evaporation, with a mass loss rate that is consistent with detailed $N$-body simulations.

The total duration of the slow-disruption phase is\footnote{See Figure~\ref{fig:tschem} for the definitions of these time-scales.} $t_{\rm halo}\equiv t_{\rm now}-t_{\rm form}-t_{\rm disc}$, which for GC formation at $z\sim3$ (i.e.~$t_{\rm now}-t_{\rm form}=11.7~\gyr$) and typical merger time-scales $t_{\rm merge}-t_{\rm form}=t_{\rm disc}<4~\gyr$ results in durations at least twice as long as for the rapid-disruption phase ($t_{\rm halo}/t_{\rm disc}>2$). Despite the fact that GCs thus spend most of their lives in the host galaxy halo, the disruption rate integrated over this phase is small (compare the characteristic values of $t_{\rm 5,cce}$ and $t_{\rm 5,evap}$ in \S\ref{sec:shake},  \S\ref{sec:stir}, and \S\ref{sec:example} below, with $t_{\rm 5,cce}/t_{\rm 5,evap}\la0.05$). As a result, the long-term survival of clusters becomes much more likely by migrating into the halo. Recall that in \S\ref{sec:intro}, we defined GCs as a bound stellar cluster that does not coincide in phase space with the presently star-forming component of the host galaxy. We now see that the migration of GC-like YMCs away from the gas-rich host galaxy disc indeed marks the moment at which they become long-lived GCs.

There exist several recent models that connect the origin of GCs to galaxy formation \citep[e.g.][]{prieto08,boley09,tonini13,katz14,li14}. However, the model presented here differs from these previous works by introducing two crucial new elements:
\begin{enumerate}
\item
It is the first model that does {\it not} make the assumption that the (globular) cluster formation rate simply scales with the SFR (or that the mass in GCs is proportional to the halo mass), but instead introduces an environmentally-dependent fraction of star formation occurring in bound stellar clusters, as well as an environmentally-dependent ICMF truncation mass. Both of these are in agreement with theory and observations of cluster populations in the nearby Universe, and the model applies them to the conditions of star formation and the ISM seen in star-forming disc galaxies at high redshift.
\item
The model is also the first to include the rapid-disruption phase in the host galaxy disc due to tidal perturbations from the ISM, in a way that is consistent with $N$-body simulations of tidally-shocked clusters. It will be shown in \S\ref{sec:comp} and \S\ref{sec:pred} that this process dominates the total mass loss of the cosmic GC population integrated over a Hubble time. As a result, the evaporation phase is sub-dominant, even though it has traditionally been the focus of GC formation models and $N$-body simulations of dissolving clusters. Effectively, these models thereby immediately place GCs in galaxy haloes and skip the (dominant) physics of GC formation and early evolution.
\end{enumerate}

Next to the addition of a detailed model for the formation and early evolution of GCs in high-redshift discs, our model also treats mass loss by evaporation differently from earlier work \citep[e.g.][]{gnedin97,prieto08,katz14,li14}. These previous studies adopted the expression from \citet{spitzer87}, which neglects the non-zero escape time of unbound stars \citep{baumgardt01} and assumes that evaporation takes place at a rate that is independent of the galactic environment. By construction, this has allowed the above previous works to reproduce the near-universal properties of GC populations (e.g.~their ÒuniversalÓ characteristic mass scale, see e.g.~\citealt{fall01}). However, this is inconsistent with numerical simulations of dissolving clusters \citep[e.g.][]{baumgardt03,gieles08,lamers10}, which show that the mass loss rate by two-body relaxation is proportional to the square root of the tidal field strength \citep{kruijssen11} or inversely proportional to the galactocentric radius (for a flat rotation curve).

In summary, the model presented here accounts for the key physical mechanisms which are known to be important for cluster formation and evolution. As such, it improves on previous work in terms of the cluster modelling. However, the coupling to galaxy formation in this (largely analytical) model is more simplified, as it makes use of a simple merger time-scale and a step function to describe the migration of clusters from discs to haloes. While this does allow us to easily identify which physical component is responsible for which aspect of the observables, the obvious benefit of previous numerical simulations is that they provide predictions for the spatial distribution and kinematics of GC populations. We will link our new model to such methods in several future papers.

\section{Comparison to observed globular cluster systems} \label{sec:comp}
The two-phase model for GC formation presented above is consistent with the formation of YMCs in the local Universe, as well as with the star-forming properties of high-redshift galaxies and the physics of hierarchical galaxy formation. The obvious next question is whether the resulting properties of the GC population agree with what is observed. We will now present the results for five different galaxy models, focusing on several of the observables discussed in \S\ref{sec:gc}. In particular, we will discuss the GC mass function (\S\ref{sec:gcmf2}), the specific frequency and metallicity bimodality of GCs (\S\ref{sec:tn2}), the relation between the GC population mass and the dark matter halo mass (\S\ref{sec:mh}), and the relation between the number of GCs  and the supermassive black hole mass (\S\ref{sec:smbh}).

\subsection{Considered example cluster population models} \label{sec:example}
\begin{table*}
 \centering
  \begin{minipage}{120mm}
  \caption{Defined ($^\star$) and derived properties of the galaxy models at $z=3$}\label{tab:init}
  \begin{tabular}{@{}l c c c c@{}}
  \hline
     & Fornax & $\feh=-1.1$ & $\feh=-0.7$ & $\feh=-0.6$ \\
  \hline
  $\log{(M_{\rm h}/\msun)}$ & $10.0^\star$ & $11.3^\star$ & $12.5^\star$ & $13.5^\star$ \\
  $\log{(M_{\rm star}/\msun)}$ & $6.6$ & $8.9$ & $10.7$ & $11.1$ \\
  $\feh$ & $-2.3^\star$ & $-1.1$ & $-0.7$ & $-0.6$ \\
  $t_{\rm merge}~[\gyr]$ & $\{1.4,0.1^\star\}$ & $2.0$ & $2.3$ & $3.4$ \\
  $\log{(M_{\rm c}/\msun)}$ & $5.8^\star$ & $5.7$ & $6.0$ & $6.1$ \\
  $\log{(M_{\rm T}/\msun)}$ & $7.6$ & $7.3$ & $7.6$ & $7.6$ \\
  $P/k~[10^6~{\rm K}~\cmc]$ & $1.06$ & $2.09$ & $6.54$ & $8.78$ \\
  $\Sigma~[\msun~\pc^{-2}]$ & $103$ & $145$ & $257$ & $297$ \\
  $\Omega~[\myr^{-1}]$ & $0.06$ & $0.09$ & $0.12$ & $0.13$ \\
  $\rho_{\rm ISM}~[\msun~\pc^{-3}]$ & $0.82$ & $1.75$ & $3.21$ & $3.70$ \\
  $f_\Sigma$ & $4.10$ & $4.01$ & $3.95$ & $3.94$ \\
  $\Gamma$ & $0.38$ & $0.46$ & $0.55$ & $0.56$ \\
  $t_{\rm cce}(10^5~\msun)~[\gyr]$ & $2.52$ & $0.36$ & $0.09$ & $0.06$ \\
  $t_{\rm evap}(10^5~\msun)~[\gyr]$ & $94.87$ & $15.81^\star$ & $2.53^\star$ & $1.26^\star$ \\
  \hline
\end{tabular}
\end{minipage}
\end{table*}
The above model is used to describe the formation and evolution of GCs in three example galaxies with $z=3$ halo masses of $M_{\rm h}=\{10^{11.3},10^{12.5},10^{13.5}\}$, for which \citet{genel09} provide merger rates. In addition, two models are included to address the extremely high mass ratio between GCs and field stars in dwarf galaxies like the Fornax dSph galaxy at metallicities $\feh<-2$ (see e.g.~\citealt{larsen12} and \S\ref{sec:gc}). For these models, the $z=3$ halo mass is taken to be $M_{\rm h}=10^{10}~\msun$, corresponding to a stellar mass of $4\times10^6~\msun$ \citep[obtained by extrapolating the galaxy mass--halo mass relation from][]{moster13}. This is roughly consistent with the total stellar mass of Fornax at metallicities $\feh<-2$ \citep{larsen12}. The mean metallicity is taken to be $\feh=-2.3$, and we adopt a ICMF truncation mass of $M_{\rm c}=7\times10^5~\msun$. The disruption time-scale during the halo phase is chosen to be $t_{\rm 5,evap}\sim10^{11}~\yr$, consistent with a galactocentric radius of $R=2~\kpc$ \citep[cf.][]{mackey03c} in the observed gravitational potential \citep{walker11}.

The two Fornax models differ only in one parameter. The merger time-scale of the first Fornax model is obtained by extrapolating a power-law fit to the \citet{genel09} relation between $t_{\rm merge}$ and $M_{\rm h}$ down to a halo mass of $M_{\rm h}=10^{10}~\msun$, resulting in $t_{\rm merge}=1.4~\gyr$. The merger time-scale of the second Fornax model is chosen to reproduce the high observed GC-to-field star mass ratio by limiting the duration of the rapid-disruption phase in the host galaxy disc, resulting in $t_{\rm merge}=0.1~\gyr$. This corresponds to a couple of orbital times. In the context of our model, such a short time-scale represents a scenario in which the high specific frequency at low metallicities in Fornax may have been caused by e.g.~the formation of the GCs during the first pericentre passage of the merger that eventually triggered their migration into the halo (see below).

Table~\ref{tab:init} shows the defined properties (indicated by an asterisk) of the galaxy models discussed in this section, as well as those derived using the physical descriptions or fitting functions described above. Having initialised the cluster population in a way consistent with observations of stellar cluster formation in the local Universe, the key question now is whether the derived galaxy properties are consistent with the `normal', star-forming disc galaxies observed at redshift $z>2$ \citep[e.g.][]{tacconi10}. The gas surface densities in Table~\ref{tab:init} place these galaxies at the low end of the surface density range of gas-rich, normal galaxies at $z>2$ \citep{tacconi13}, indicating that the conditions needed for GC formation are indeed common at high redshift. The corresponding pressures are also modest, falling in the range $P/k=10^6$--$10^7~{\rm K}~\cmc$. Although this is 2--3 orders of magnitude higher than the turbulent and hydrostatic pressures in the Milky Way disc, it is within the range of pressures observed at $z>2$, which reach maxima of $P/k=10^8$--$10^9~{\rm K}~\cmc$ \citep[see \S\ref{sec:hiz} and e.g.][]{swinbank12}. The resulting gas volume densities are up to 100 times higher than in the Milky Way disc, but again consistent with the prevalent high-density gas tracer emission detected in high-redshift galaxies \citep[e.g.~from high-$J$ CO transitions, which suggest the presence of large mass reservoirs with {\it peak} densities $\rho>50~\msun~\pc^{-3}$, see][]{danielson13}. As Table~\ref{tab:init} shows, the increase of the gas pressure is accompanied by a corresponding increase of the fraction of star formation occurring in bound clusters $\Gamma$, as well as a strong decrease of the cluster disruption time-scale $t_{\rm cce}$.

Of course, observed $z=0$ galaxies will host a composite GC population with a range of metallicities and formation redshifts, rather than the `Simple GC Populations' of Table~\ref{tab:init}, which each have a single age and metallicity. The galaxy models presented here are chosen to cover the GC metallicity range. If GCs are indeed the products of regular YMC formation at high redshift, these galaxy models should reproduce the observed properties of the present-day GC population as a function of metallicity. In the model, we account for the further growth and composite nature of GC populations by carrying out a summation over a set of models with different metallicities (see \S\ref{sec:stir} and \S\ref{sec:gcmf2}).

\subsection{Two-phase cluster disruption and the globular cluster mass function} \label{sec:gcmf2}
\begin{figure*}
\center\includegraphics[width=\hsize]{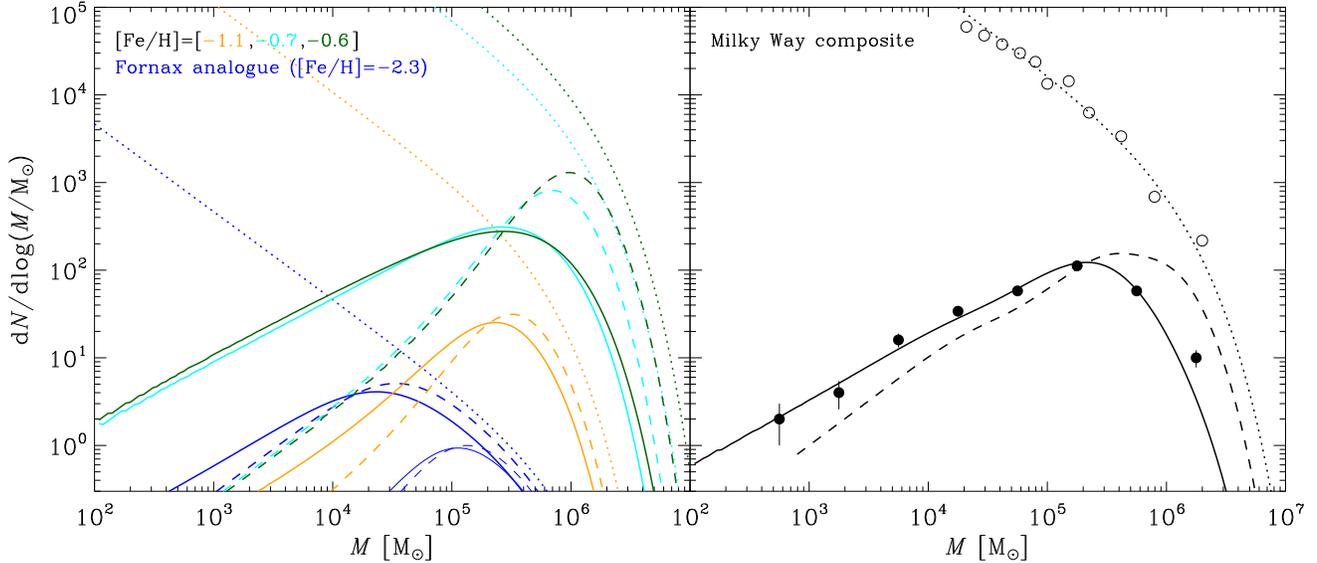}
\caption{This figure shows that the combination of regular cluster formation at $z>2$ with subsequent, two-phase disruption yields a peaked GCMF with a near-universal peak mass, consistent with observations of present-day GC populations. {\it Left panel}: Shown is the number of GCs per unit $\log{M}$ for the five different galaxies from Table~\ref{tab:init}. The legend indicates the colour coding, where the \{thin,thick\} blue lines represent the \{long,short\}-$t_{\rm merge}$ Fornax model (towards the bottom right and top left in the figure, respectively). Dotted lines show the initial cluster mass function at formation ($t_{\rm form}$), dashed lines show the cluster mass function at the end of the rapid-disruption phase ($t_{\rm merge}$), and solid lines show the GC mass function at $z=0$ ($t_{\rm now}$). The peak mass reaches a maximum at $t_{\rm merge}$ due to the destruction of low-mass clusters in the host galaxy disc, after which it decreases due to the further evaporation of the massive, surviving GCs. {\it Right panel}: Shown is the number of GCs per unit $\log{M}$ for a composite-metallicity GC population representative for the Milky Way. The black curves are obtained by a linear combination of the single-metallicity and single-age GC populations shown by the coloured lines in the left-hand panel, scaling each single-metallicity GCMF by the number of Galactic GCs in the corresponding metallicity bins (see the text). At the lowest metallicities, the mean GCMF of the two Fornax models was used. The filled dots with Poisson error bars show the observed GCMF of the Milky Way \citep[2010 edition]{harris96}, whereas open dots show the ICMF of YMCs in the Antennae galaxies \citep[multiplied by a factor of 30]{zhang99}. Both provide a good match to the modelled cluster mass functions, illustrating how observed GCMFs can be reproduced by a linear combination of the single-metallicity model GCMFs.} \label{fig:gcmf}
\end{figure*}

The left-hand panel of Figure~\ref{fig:gcmf} shows the shape of the cluster mass function for the five galaxy models of Table~\ref{tab:init} at the three key moments of cluster formation, their migration, and the present day ($t_{\rm form}$, $t_{\rm merge}$, and $t_{\rm now}$ as defined in Figure~\ref{fig:tschem} -- note that $t_{\rm merge}$ is different for each of the models because it increases with halo mass and metallicity). The figure shows that the Schechter-type ICMFs with slightly different truncation masses $M_{\rm c}$ evolve to a peaked shape during the rapid-disruption phase in the host galaxy disc, due to the tidal disruption of low-mass clusters. The peak mass varies substantially with the host galaxy, but is almost universally larger than the present day peak mass (i.e.~$M_{\rm peak}(t_{\rm merge})>M_{\rm peak}(t_{\rm now})=2\times10^5~\msun$). The possible increase of the peak mass to even higher masses is arrested by the decrease of the dynamical time with cluster density (and with mass, given that $r_{\rm h}\propto M^\beta$ with $\beta<1/3$, see \S\ref{sec:ymc}), which for clusters beyond the peak mass becomes so short that they can respond adiabatically to tidal perturbations and do not suffer the violent mass loss that destroys the lower-mass clusters. This transition from impulsive perturbations to adiabatic expansion causes a slight steepening of the dashed curves at $M\sim10^5~\msun$. In addition, the further increase of the peak mass due to disruption is also decelerated because it runs into the exponential truncation of the ICMF \citep[cf.][]{gieles09}.

During the subsequent slow-disruption phase in the galaxy halo, the peak mass decreases due to the gradual, evaporation-driven mass loss of the clusters that survived long enough to migrate into the halo. As discussed in \S\ref{sec:stir}, we have assumed a mass loss rate due to evaporation such that the observed $M_{\rm peak}(t_{\rm now})\sim2\times10^5~\msun$ is reproduced. At a time $t_{\rm merge}$ (dashed lines in Figure~\ref{fig:gcmf}), higher-metallicity models have higher peak masses, hence our assumption implies that high-metallicity GCs must lose more mass due to evaporation, i.e.~the corresponding disruption time-scale $t_{\rm 5,evap}$ must decrease with metallicity to move the peak masses back to a similar mass-scale by $z=0$. We discussed in \S\ref{sec:stir} that the resulting adopted relation of a decreasing $t_{\rm 5,evap}\propto\Omega^{-1}\propto R$ with $\feh$ is quantitatively consistent with the observed metallicity gradient of the Galactic GC population. This implies that even though $t_{\rm 5,evap}$ was tuned to reproduce the observed $M_{\rm peak}$, the result is equivalent to adopting the theoretically predicted evaporation time-scales. This is an important test of the model.

The above results show that our model predicts a near-universal peak mass of the GCMF at $z=0$. The rapid destruction of low-mass clusters in their host galaxy discs is most effective at high galaxy masses and metallicities, causing the universally-high $M_{\rm peak}$ to increase with metallicity by the end of the rapid-disruption phase (i.e.~at $t_{\rm merge}$). This $M_{\rm peak}$--$\feh$ relation is subsequently washed out by the opposite metallicity-dependence of mass loss due to evaporation in the galaxy halo, in which metal-rich GCs reside at smaller galactocentric radii and lose more mass than metal-poor GCs, resulting in a near-universal peak mass after a Hubble time. As discussed in \S\ref{sec:gradient}, the assumption underpinning this prediction is that the relation between $\feh$ and $t_{\rm 5,evap}\propto\Omega^{-1}\propto R$ observed in the Milky Way also holds in other galaxies to within a factor of several. This should be expected if the relation reflects a more fundamental relation between the GC metallicity and the binding energy within the host galaxy at the time of GC formation, which in turn would simply reflect the galaxy mass-metallicity relation.

In our model, the GC population is divided into metallicity bins to enable a direct connection to the properties of the host galaxy at the time of GC formation through the galaxy mass-metallicity relation. However, the GC populations of observed galaxies are constituted by a broad range of metallicities (see \S\ref{sec:gc}). The right-hand panel of Figure~\ref{fig:gcmf} shows the evolution from the Schechter-type ICMF at $z=3$ to the peaked GCMF at $z=0$ for a composite-metallicity GC population representative for the Milky Way. This panel accounts for the hierarchical growth of the metallicity-composite Galactic GC population by combining the GCMFs for the different metallicities shown in the left-hand panel, each scaled by the number of Galactic GCs in the corresponding metallicity bin (using the 2010 edition of the \citealt{harris96} catalogue, with bin separations at $\feh=\{-1.7,-0.9,-0.65\}$). The figure shows that at $z=3$, our model matches the ICMF of YMCs observed in the Antennae galaxies, whereas it reproduces both the normalisation and shape of the GCMF of the Milky Way at $z=0$ (note that this model has {\it not} been fit to reproduce the data).\footnote{The slight discrepancy at $M>10^6~\msun$ may exist because the model does not include the former nuclear clusters of tidally stripped dwarf galaxies (see \S\ref{sec:open}), of which there may be a handful of examples in the Galactic halo (such as $\omega$Cen, see e.g.~\citealt{lee09}). However, the difference is also consistent with the uncertainties on the model parameters.} The near-universal peak mass was already evident when considering the different metallicities individually, but for composite-metallicity GC populations any remaining, small variations average out and the universality of the peak mass becomes almost inescapable.

The two-phase evolution of the GCMF contrasts strongly with the classical idea that the evolution from the power-law ICMF to a peaked GCMF is caused by evaporation only. In that model, haloes with low GC evaporation rates should have low peak masses. In the two-phase model, there exist no GC systems with low peak masses because most of the evolution towards a peaked GCMF took place in the first few $\gyr$ after the clusters were formed. In our model, the approximate universality of the peak mass disappears only at metallicities where the present-day peak mass is not attained during the rapid-disruption phase [i.e.~$M_{\rm peak}(t_{\rm merge})<2\times10^5~\msun$]. We find this occurs at metallicities of $\feh\la-1.5$. Towards lower metallicities (and hence galaxy masses), we should thus expect a trend of decreasing $M_{\rm peak}$. This is remarkably consistent with the observed, weak trend between $M_{\rm peak}$ and GC $\feh$, which sets in at $M_B\ga-17$ and hence $\feh\la-1.3$ (as seen by combining Figure~13 of \citealt{peng06} with Figure~15 of \citealt{jordan07}).

As shown by the thick blue solid line, the two-phase model also allows for the existence of Fornax-like galaxies with a high number of GCs per unit stellar mass at $\feh<-2$, provided that $t_{\rm disc}\equiv t_{\rm merge}-t_{\rm form}$ was short. In this model, the host galaxy experienced a merger some $10^8~\yr$ after the formation of the young GCs. Because the densest and most massive clusters can respond adiabatically to tidal perturbations, only clusters below a critical density or mass (as $r_{\rm h}\propto M^\beta$ with $\beta<1/3$, see \S\ref{sec:ymc}) get destroyed, whereas clusters above that critical mass hardly lose any stars. For our short-$t_{\rm merge}$ Fornax model, this transition occurs at $M\sim10^{4.3}~\msun$.

While a short migration time-scale ensures the survival of a larger number of GCs than for normal migration time-scales, the shape of the observed GCMF in Fornax is not fully consistent with that shown here. The observed peak mass is higher than in the model by a factor of $\sim3$. This could potentially be caused by dynamical friction -- if the present-day orbits of the GCs (which were used to determine $t_{\rm 5,evap}$) were wider in the past, the total amount of mass loss during the halo phase may have been minimal. However, the low peak mass is already present at $t_{\rm merge}$, suggesting that the discrepancy already arises during the rapid-disruption phase. It could be that we underestimated the effect of disruption. Alternatively, the short-$t_{\rm merge}$ model was designed to match the high observed specific frequency of Fornax (see below), but did so while keeping the minimum cluster mass constant. If the minimum cluster mass was higher in Fornax, the observed specific frequency would require more cluster disruption (i.e.~a later migration of the GCs into the halo), which would increase the peak mass.

Finally, we note that the difference between the GCMFs of the two Fornax models illustrates the maximum impact of the stochasticity of the merger process. The adopted time-scale for the long-$t_{\rm merge}$ model (thin blue lines in Figure~\ref{fig:gcmf}) represents a cosmic average, but since galaxy merging is a Poisson process, deviations to much shorter (thick blue lines in Figure~\ref{fig:gcmf}) or longer merger time-scales (and hence $t_{\rm disc}$) are possible. The importance of Poisson noise should decrease with galaxy mass, because the GC populations of more massive galaxies should originate from a larger variety of progenitor galaxies.

\subsection{The specific frequency as a function of metallicity and stellar mass} \label{sec:tn2}
\begin{figure}
\center\includegraphics[width=\hsize]{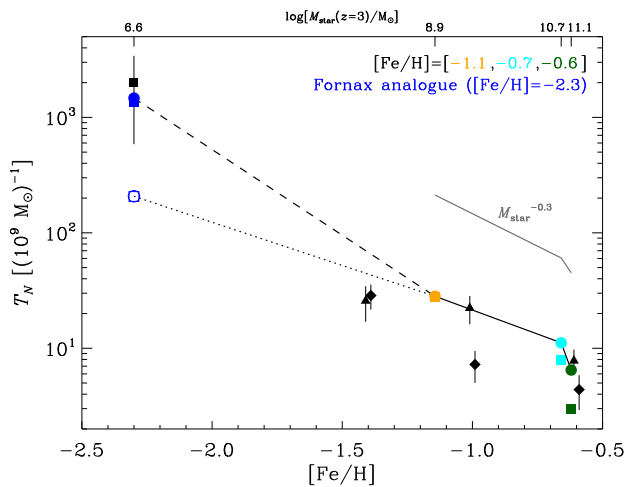}
\caption{This figure shows that the GC specific frequency decreases with host galaxy mass and metallicity, because the destruction of clusters during the rapid-disruption phase in their host galaxy discs both proceeds more rapidly and lasts longer in massive galaxies. Shown is the specific frequency $T_N\equiv N_{\rm GC}(M_{\rm star}/10^9~\msun)^{-1}$ as a function of the host galaxy metallicity (bottom x-axis) and stellar mass (top x-axis), for the five different galaxies from Table~\ref{tab:init}. The legend indicates the colour coding of the coloured symbols, where the \{open,filled\} blue symbols represent the \{long,short\}-$t_{\rm merge}$ Fornax model. Coloured circles show the specific frequencies at the end of the rapid-disruption phase ($t_{\rm merge}$), whereas coloured squares show the specific frequencies at $z=0$ ($t_{\rm now}$). The grey line indicates the analytically predicted relation $T_N\propto M_{\rm star}^{-0.3}$ (see the text). The figure shows that the relation of decreasing $T_N$ with $\feh$ and $M_{\rm star}$ is established at $t_{\rm merge}$ due to rapid cluster disruption, after which the number of GCs remains roughly constant except for the most metal-rich GCs. The lack of evolution during the halo phase is consistent with the conclusion of \S\ref{sec:gc} that the specific frequency of GCs does not depend on the galactocentric radius and hence must have been set during their formation or early evolution. For comparison, we again include the GC populations of NGC~5128 and Fornax as black symbols (see Figure~\ref{fig:tnfeh_obs}).} \label{fig:tnfeh}
\end{figure}

Figure~\ref{fig:tnfeh} shows the number of GCs per unit galaxy stellar mass [the specific frequency $T_N\equiv N_{\rm GC}(M_{\rm star}/10^9~\msun)^{-1}$] as a function of the host galaxy metallicity and stellar mass for the five galaxy models of Table~\ref{tab:init} at the time of GC migration and at the present day ($t_{\rm merge}$ and $t_{\rm now}$, see Figure~\ref{fig:tschem}). The figure shows that the specific frequency decreases with the metallicity and host galaxy mass, which is caused by an increase of both the disruption rate and merger time-scale (i.e.~the duration of the rapid-disruption phase) with galaxy mass (see Table~\ref{tab:init}). We find that this decrease is quantitatively consistent with the observed trend (also see below).

In \S\ref{sec:gc}, we have drawn particular attention to the relation between $T_N$ and the host galaxy mass at the time of GC formation $t_{\rm form}$, because its observed invariance with galactocentric radius implies that it cannot be affected by GC evaporation in the galaxy halo. It must therefore have been set shortly after GC formation. Figure~\ref{fig:tnfeh} shows that this is indeed the case -- the specific frequencies are in place at $t_{\rm merge}$ and evolve very little during the halo phase until $t_{\rm now}$. This is consistent with the result shown in Figure~\ref{fig:gcmf} above, where we discussed how the {\it number of clusters} decreases by tidal destruction during the rapid-disruption phase, whereas during the halo phase the {\it masses} of the surviving GCs decrease and few GCs are being destroyed altogether. Consequently, $N_{\rm GC}$ and $T_N$ are set at $t_{\rm merge}$ (i.e.~at $z=1$--$2$), whereas the GC masses and radii undergo some mild further evolution within the host galaxy haloes.

It can be shown analytically how the relation between the specific frequency and the host galaxy mass and metallicity is shaped by cluster disruption. The specific frequency represents the number of surviving GCs per unit stellar mass and hence combines a large part of all the cluster formation and disruption processes discussed in \S\ref{sec:phys}. However, the number of YMCs (i.e.~young GCs) initially formed per unit stellar mass shows little variation with galaxy mass (see Table~\ref{tab:init}), because it only depends on (the weakly galaxy mass-dependent) $\Gamma$ and $M_{\rm c}$. The decrease of the specific frequency with galaxy mass must therefore reflect an increasing trend of the amount of cluster disruption (i.e.~a decreasing survival fraction). Table~\ref{tab:init} shows that the rate and duration of cluster disruption during the rapid-disruption phase in the host galaxy disc does indeed exhibit a strong dependence on the host galaxy mass. The disruption time-scale due to tidal shocks $t_{\rm cce}$ is expressed in equation~(\ref{eq:tcce}), which reflects {\it how rapidly} clusters are being disrupted, whereas the migration (i.e.~merger) time-scale $t_{\rm merge}$ reflects {\it how long} clusters are being disrupted. The ratio between these disruption and merger time-scales thus probes the survival fraction, i.e.~$T_N\propto t_{\rm cce}/t_{\rm merge}$. The tidal disruption time-scale scales with the host galaxy's stellar mass at $z=3$ as $t_{\rm cce}\propto\rho_{\rm ISM}^{-3/2}\propto M_{\rm star}^{-0.22}$ (see \S\ref{sec:init}), whereas the merger time-scales of Table~\ref{tab:init} satisfy a power-law relation of $t_{\rm merge}=2.4~\gyr~(M_{\rm star}/10^{10}~\msun)^{0.084}$. As a result, the specific frequency should decrease with stellar mass as $T_N\propto M_{\rm star}^{-0.3}$, which is indeed seen in Figure~\ref{fig:tnfeh}.

The above line of reasoning shows that the fundamental relation of Figure~\ref{fig:tnfeh} is one between $T_N$ and the host galaxy mass at the time of GC formation, because both the gas density and the merger rate primarily depend on the galaxy mass in our model. However, the information on the initial host galaxy mass is lost during the subsequent growth and merging of the galaxies hosting GCs. To still trace the mass-dependence of the specific frequency in the local Universe, the galaxy mass-metallicity relation at the formation redshift of GCs can be used to translate the $T_N$-$M_{\rm star}$ relation to a correlation between $T_N$ and $\feh$. This explains the origin of the observed $T_N$--$\feh$ relation.

For metallicity-composite GC populations, the monotonic decrease of $T_N$ with galaxy mass and metallicity shown in Figure~\ref{fig:tnfeh} should be reversed at high galaxy masses through hierarchical galaxy growth. Massive, metal-rich galaxies accrete a large fraction of their mass ($20$--$50\%$ for $M_{\rm h}=10^{13}$--$10^{14}~\msun$, see \citealt{moster13}) in the form of low-metallicity galaxies with high $T_N$, implying that the specific frequency of the massive galaxy must increase. Because the accretion rate of low-mass galaxies increases with galaxy mass, this can lead to a minimum of the $z=0$ relation between the specific frequency and the stellar mass. The galaxy mass at which this minimum is reached corresponds to the point where the GC population becomes dominated by the metal-poor GCs accreted from stripped dwarf galaxies. Figure~8 of \citet{peng08} shows that this is the case for $M_B<-22$, which indeed roughly corresponds to the upturn of the specific frequency towards high galaxy masses \citep[top-left panel of Figure~6]{peng08}. For such a composite population of GCs with different metallicities (and possibly formation redshifts), the galaxy-wide specific frequency can be obtained by integrating the $T_N$--$\feh$ relation of Figure~\ref{fig:tnfeh} over the halo field star metallicity distribution function. This integration should account for the halo star formation history and the dependence of $T_N$ on the formation redshift. For instance, the survival fractions of newly-formed clusters decrease strongly with cosmic time due to the decreasing galaxy merger rates and Toomre masses, whereas the stellar mass may continue to grow by in-situ star formation.

Analogously to the GCMF in \S\ref{sec:gcmf2}, the $T_N$ difference between the two Fornax models illustrates the maximum impact of the stochasticity of the merger process. While the long-$t_{\rm merge}$ model yields a specific frequency of $T_N\sim200$, the short-$t_{\rm merge}$ model reproduces the observed specific frequency of $T_N\sim2\times10^3$. This order-of-magnitude difference agrees well with the scatter of the specific frequency at low galaxy masses \citep[e.g.][]{peng08,georgiev10} and should decrease with galaxy mass due to the increasing importance of (more numerous and hence less stochastic) minor mergers. Despite the model's good agreement with the Fornax dSph, Figure~\ref{fig:gcmf} shows that the mean GC mass in the model is a factor of $2$--$3$ lower than observed in Fornax ($M\sim10^5~\msun$). As a result, the {\it number} of GCs per unit stellar mass is well-reproduced by the model, but GCs constitute only 9\% of all coeval stellar mass at $z=0$, about a factor of two lower than the observed percentage \citep{larsen12}. None the less, the model does quite well in describing the statistics of Fornax' GC population, showing that a short migration time-scale is a feasible hypothesis for explaining the high specific frequency at $\feh<-2$ in the Fornax dSph.

In \S\ref{sec:gc}, it was posited that the $T_N$--$M_{\rm star}(t_{\rm form})$ relation of Figure~\ref{fig:tnfeh} is of fundamental importance to a theory of GC formation, because it implies several of the other characteristics of GC populations. For instance, \citet{tonini13} used the $z=0$ relation between $T_N$ and $M_{\rm star}$ (which is a metallicity and formation redshift composite of the specific frequencies shown in Figure~\ref{fig:tnfeh}) to initialise GC populations at redshifts $z=2$--$4$ and show that hierarchical galaxy formation naturally produces the observed colour (and hence metallicity) bimodality observed in the GC populations of massive ($M_{\rm star}\ga10^9~\msun$) galaxies (see \S\ref{sec:gc}). The \citet{tonini13} model works, because the specific frequency increases steeply towards low galaxy masses and metallicities, which means that accreted dwarf galaxies can boost the number of low-$\feh$ GCs even if their impact on the field star metallicity distribution is negligible. However, this argument relies on the ad-hoc assumption that the $T_N$--$M_{\rm star}$ relation was already in place at $z\sim2$. Figure~\ref{fig:tnfeh} shows that this is indeed the case, because the GC destruction that sets the $T_N$--$M_{\rm star}$ relation takes place before the GCs have migrated into the halo. Combining the physical model for the origin of the $T_N$--$M_{\rm star}$ relation presented here with the result from \citet{tonini13}, we have thus captured the physics behind the bimodality of GC colour and metallicity observed at $z=0$.

\subsection{The globular cluster system mass per unit halo mass} \label{sec:mh}
\begin{figure}
\center\includegraphics[width=\hsize]{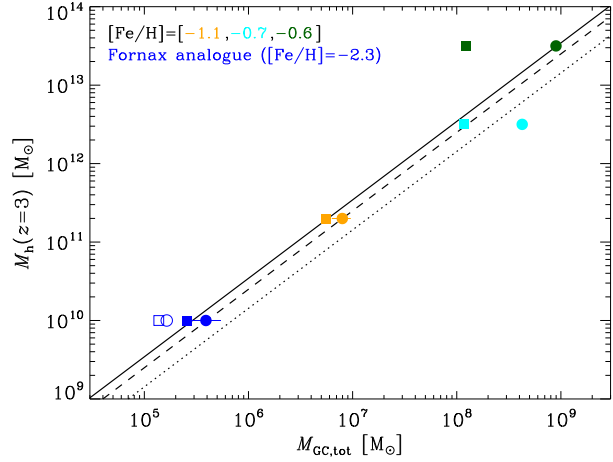}
\caption{This figure shows that the total GC system mass per unit host galaxy halo mass is constant. Shown is the dark matter halo mass as a function of the total mass of the GC population, for the five different galaxies from Table~\ref{tab:init}. The legend indicates the colour coding, where the \{open,filled\} blue symbols represent the \{long,short\}-$t_{\rm merge}$ Fornax model. Coloured circles show the halo and total GC masses at the end of the rapid-disruption phase ($t_{\rm merge}$), whereas coloured squares show the halo and total GC masses at $z=0$ ($t_{\rm now}$). The solid line shows the observed relation from the Next Generation Virgo Cluster Survey \citep{durrell14}, which covers galaxy haloes out to the virial radius. Other observational estimates are shown as the dashed \citep{hudson14} and dotted \citep{spitler09} lines. The figure shows that there is little evolution of $\eta\equiv M_{\rm GC,tot}/M_{\rm h}$ between $t_{\rm merge}$ and $t_{\rm now}$, consistent with the similar lack of evolution shown in Figures~\ref{fig:gcmf} and~\ref{fig:tnfeh}. Only for the most metal-rich GCs there is a deviation, but in observed $z=0$ galaxies this effect is washed out because most GCs have metallicities $\feh<-0.6$, for which this figure shows that $\eta\sim{\rm constant}$. As is discussed in the text, the constant GC system mass per unit halo mass arises because the decrease of the specific frequency with galaxy mass (see Figure~\ref{fig:tnfeh}) is cancelled by the increase of the stellar mass fraction with galaxy mass.} \label{fig:mgcmh}
\end{figure}
Figure~\ref{fig:mgcmh} shows the relation between the dark matter halo mass $M_{\rm h}$ and the total mass of the galaxy's GC population $M_{\rm GC,tot}$ for the five galaxy models of Table~\ref{tab:init} at the time of GC migration and at the present day ($t_{\rm merge}$ and $t_{\rm now}$, see Figure~\ref{fig:tschem}). As discussed in \S\ref{sec:gc}, this relation has received considerable attention in the recent observational literature. The near-linear slope (i.e.~the constant ratio $\log{\eta}\equiv M_{\rm GC,tot}/M_{\rm h}=-4.5$, \citealt{durrell14}) is quite suggestive of a common origin of GCs and dark matter haloes \citep[e.g.][]{spitler09}. However, Figure~\ref{fig:mgcmh} shows that the model presented here also gives rise to a constant GC system mass per unit halo mass, even though the GC formation rate is not assumed to scale with the host halo mass. It will be shown below that this relation follows naturally from the combination of GC disruption (through the $T_N$--$M_{\rm star}(t_{\rm form})$ relation from Figure~\ref{fig:tnfeh}) and the physics of galaxy formation within dark matter haloes.

The modelled $M_{\rm GC,tot}$--$M_{\rm h}$ relation of Figure~\ref{fig:mgcmh} agrees well with the observed relations \citep{spitler09,hudson14,durrell14}, with a normalisation that is similar to within a factor of two. The only discrepancy exists for the $\feh=-0.6$ galaxy, but this should disappear in the composite GC populations of $z=0$ galaxies, because they are constituted by a mix of different metallicities, and therefore contain only a small fraction of $\feh>-0.6$ GCs. The difference between the latest observations \citep{hudson14,durrell14} and the earlier relation from \citet{spitler09} is caused by a difference in spatial coverage. The Next Generation Virgo Cluster Survey \citep{ferrarese12} as well as the lensing results from \citet{hudson14} now cover a large sample of galaxies all the way out to their virial radii, implying a slight increase of the dark matter mass per unit GC mass relative to \citet{spitler09}. This upward shift is in agreement with the model results.

We can construct a simple metallicity-composite GC population by assuming that $\sim1/6$ of the GCs has $\feh>-0.6$ \citep[e.g.][]{harris96,peng06}. This results in predicted values of $\log{\eta}=-4.2$ at $t_{\rm merge}$ and $\log{\eta}=-4.6$ at $t_{\rm now}$, which are both within a factor of 2--3 of all observed GC population-to-halo mass relations. This is remarkable for two reasons. Firstly, we have not assumed any intrinsic connection between GCs and dark matter haloes (such as the idea that GC formation took place during reionization as the first haloes collapsed). In the model, GCs are the products of regular star and cluster formation in high-redshift galaxies, yet a one-to-one relation between GCs and haloes emerges naturally. Secondly, the model is very simple and omits e.g.~the composite formation redshifts of $z=0$ GC populations and further galaxy growth between $t_{\rm merge}$ and $t_{\rm now}$.  The good agreement between theory and observations suggests that any mix between GC sub-populations at different ages and metallicities should retain the constant value of $\eta$. Indeed, the above example of a metallicity-composite GC population illustrates how hierarchical galaxy growth will further decrease the scatter of $\eta$ across the galaxy population due to the central limit theorem.

The obvious question is what causes the ratio $M_{\rm GC,tot}/M_{\rm h}$ to be constant. The model allows us to identify the reason by expressing $\eta$ in terms of the $T_N$--$M_{\rm star}$ relation seen in Figure~\ref{fig:tnfeh}. By writing
\begin{equation}
\label{eq:eta}
\eta\equiv\frac{M_{\rm GC,tot}}{M_{\rm h}}\propto\frac{N_{\rm GC}}{M_{\rm h}}=\frac{N_{\rm GC}}{M_{\rm star}}\frac{M_{\rm star}}{M_{\rm h}}=T_N\frac{M_{\rm star}}{M_{\rm h}} ,
\end{equation}
we see that $\eta$ is the product of $T_N\propto M_{\rm star}^{-0.3}$ (see \S\ref{sec:tn2}) and the stellar-to-halo mass ratio. For $M_{\rm h}<10^{12.5}~\msun$, \citet{moster13} show that the relation between galaxy stellar mass and halo mass at $z=3$ is $M_{\rm h}\propto M_{\rm star}^{0.6}$, resulting in $M_{\rm star}/M_{\rm h}\propto M_{\rm star}^{0.4}$. Substituting this into equation~(\ref{eq:eta}), we indeed find that $\eta$ is nearly independent of the galaxy mass with $\eta\propto M_{\rm star}^{0.1}$.

Present-day galaxies are constituted by composite stellar populations assembled through the merging of the single-age and single-metallicity populations considered here. Between $z=3$ and $z=0$, this merging process must have further diluted the already-weak correlation between $\eta$ and galaxy mass seen in the single-metallicity models of Figure~\ref{fig:mgcmh}. For halo masses $M_{\rm h}>10^{12.5}~\msun$, the stellar mass fraction decreases with the halo mass, in stark contrast with the increase at lower masses that leads to a constant value of $\eta$ through equation~(\ref{eq:eta}), indicating that the GC system mass per unit halo mass should decrease in massive haloes. However, the GC populations in massive haloes are generally observed to have mean metallicities $\feh<-0.7$ \citep{peng06}, indicating that the majority of their GCs originates from lower-mass galaxies. As a result, the ratio between the GC population mass and the halo mass is still dominated by GCs that formed in haloes with $M_{\rm h}<10^{12.5}$ for which $M_{\rm star}/M_{\rm h}\propto M_{\rm star}^{0.4}$ and hence $\eta\sim{\rm constant}$. Only for galaxy masses of $M_{\rm star}>10^{11.3}~\msun$ at $z=0$ (corresponding to $M_{\rm h}>10^{13.5}~\msun$), the mean metallicity of the GC population reaches $\feh\sim-0.7$, possibly leading to a slight steepening of the $M_{\rm h}$--$M_{\rm GC,tot}$ relation. There is a tentative hint for such a steepening in Figure~3 of \citet{spitler09}, but a larger sample of high-mass galaxies is necessary to confirm this. Such a sample should also enable the identification of second-order dependences, such as on GC sub-population metallicity and galaxy morphology \citep{harris15}, which may reflect differences in the formation histories of the host galaxy and its progenitors.

In summary, our model suggests that a constant GC system mass per unit halo mass arises naturally because both the number of surviving GCs per unit stellar mass (Figure~\ref{fig:tnfeh}) and the halo mass per unit stellar mass \citep{moster13} decrease with the host galaxy stellar mass at the same rate for halo masses $M_{\rm h}<10^{12.5}~\msun$. When taking the ratio between $M_{\rm GC,tot}$ and $M_{\rm h}$, the dependence on the stellar mass cancels, resulting in $\eta\sim{\rm constant}$. This shows that the observed scaling is a `coincidence' and can be understood without requiring special, early-Universe physics.

\subsection{The scaling between the number of globular clusters and the supermassive black hole mass} \label{sec:smbh}
Using a sample of nearby early-type galaxies, it was recently shown by \citet{burkert10} that the number of GCs scales with the mass of the host galaxy's central supermassive black hole (SMBH) $M_\bullet$, i.e.~$N_{\rm GC}\propto M_\bullet^{0.94\pm0.10}$. Follow-up work by \citet{harris11} has confirmed this result, although observations of SMBHs in spiral galaxies suggest that the relation may be significantly shallower in late-type systems (Lomeli et al.~in prep.). Irrespective of the exact slope of the $N_{\rm GC}$--$M_\bullet$ relation, it is almost certainly not causal in nature -- there is no known physical mechanism that directly connects a galaxy's GC population to the central SMBH, other than the suggestion that both classes of objects assembled mainly at early cosmic times \citep[e.g.][]{kruijssen14c,sijacki15}. None the less, we can test if our model reproduces the empirical $N_{\rm GC}$--$M_\bullet$ relation by combining the observed scaling relations between the SMBH mass and the host galaxy properties with the predicted relations between the number of GCs and the host galaxy properties.
\begin{table*}
 \centering
  \begin{minipage}{140mm}
  \caption{Properties of the GC populations at the times $t=\{t_{\rm merge},t_{\rm now}\}$.}\label{tab:comp}
  \begin{tabular}{@{}l c c c c c@{}}
  \hline
     & Fornax (1) & Fornax (2) & $\feh=-1.1$ & $\feh=-0.7$ & $\feh=-0.6$ \\
  \hline
  $M_{\rm peak}(t_{\rm merge})~[10^5~\msun]$ & $1.2$ & $0.4$ & $2.7$ & $5.7$ & $7.8$ \\
  $M_{\rm peak}(t_{\rm now})~[10^5~\msun]$ & $1.0$ & $0.3$ & $1.8$ & $2.3$ & $2.5$ \\
  $N_{\rm GC}(t_{\rm merge})$ & $1$ & $6$ & $23$ & $583$ & $902$ \\
  $N_{\rm GC}(t_{\rm now})$ & $1$ & $6$ & $23$ & $415$ & $413$ \\
  $T_N(t_{\rm merge})~[(10^9~\msun)^{-1}]$ & $207.3$ & $1468.8$ & $28.2$ & $11.2$ & $6.5$ \\
  $T_N(t_{\rm now})~[(10^9~\msun)^{-1}]$ & $206.4$ & $1354.7$ & $27.6$ & $7.9$ & $3.0$ \\
  $M_{\rm GC,tot}(t_{\rm merge})/M_{\rm star}~[10^{-2}]$ & $3.8$ & $8.9$ & $1.0$ & $0.8$ & $0.6$ \\
  $M_{\rm GC,tot}(t_{\rm now})/M_{\rm star}~[10^{-2}]$ & $3.2$ & $5.9$ & $0.7$ & $0.2$ & $0.1$ \\
  $M_{\rm GC,tot}(t_{\rm merge})/M_{\rm h}~[10^{-5}]$ & $1.6$ & $3.9$ & $4.0$ & $13.4$ & $2.8$ \\
  $M_{\rm GC,tot}(t_{\rm now})/M_{\rm h}~[10^{-5}]$ & $1.4$ & $2.6$ & $2.8$ & $3.7$ & $0.4$ \\
  \hline
\end{tabular}
\end{minipage}
\end{table*}

The SMBH mass is primarily related to the velocity dispersion or stellar mass of the host galaxy spheroid \citep{magorrian98,ferrarese00,gebhardt00}, i.e.~a specific galactic component. Unfortunately, relations specific to a single component are not suitable to test whether the $N_{\rm GC}$--$M_\bullet$ relation is consistent with the GC formation model presented here, because our model does not distinguish between the morphological components of the host galaxy. Instead, we can use the relation between the dark matter halo mass $M_{\rm h}$ and the SMBH mass of $M_{\rm h}\propto M_\bullet^{0.61}$ from \citet{ferrarese02}, which may be less fundamental than the spheroid-based relations \citep{kormendy13} but does hold empirically. For halo masses $M_{\rm h}<10^{12.5}~\msun$ at $z=0$, the stellar mass scales with the halo mass as $M_{\rm star}\propto M_{\rm h}^{2.3}$, which using the \citet{ferrarese02} relation implies $M_{\rm star}\propto M_\bullet^{1.4}$. As discussed in \S\ref{sec:tn2}, the presented model predicts $N_{\rm GC}\propto M_{\rm star}(z=3)^{0.7}$. Without a model that describes the further growth of galaxies and SMBHs between $z\sim3$ and $z=0$, there is no reason to assume that this relation can be directly connected to the observed $M_{\rm star}$--$M_\bullet$ relation at $z=0$. However, combining both expressions does yield $N_{\rm GC}\propto M_\bullet^{0.98}$, in good agreement with the observed relation. In the context of this work, the good agreement between the $N_{\rm GC}$-$M_\bullet$ relation in the $z\sim3$ model and the $z=0$ observations suggests that the SMBH mass and the host galaxy stellar mass evolved roughly in proportion to each other between $z\sim3$ and $z=0$, as is indeed found in galaxy formation simulations \citep[e.g.][]{sijacki15}.

\subsection{Summary of the presented results and comparison to previous models}
The results of this section show that the observed properties of GCs at $z=0$ are reproduced by our model, indicating that GCs are fully consistent with being the products of the regular star and cluster formation process in the high-pressure environments of $z>2$ galaxies. These clusters were saved from tidal destruction in their dense natal environments by the redistribution of matter (and hence their `liberation' into the galaxy halo) during hierarchical galaxy formation. The most important quantitative constraints from Figures~\ref{fig:gcmf}--\ref{fig:mgcmh} are summarised in Table~\ref{tab:comp}, which lists the GCMF peak mass $M_{\rm peak}$, the number of GCs $N_{\rm GC}$, the specific frequency $T_N$, the ratio between the GC population mass and host galaxy stellar mass $M_{\rm GC,tot}/M_{\rm star}$, and the ratio between the GC population mass and the host galaxy halo mass $M_{\rm GC,tot}/M_{\rm h}$, at times $t_{\rm merge}$ and $t_{\rm now}$ (as defined in Figure~\ref{fig:tschem}).

Previous models aimed at explaining the origin of GCs in the context of galaxy formation have almost all included a description for the origin of the GCMF \citep[e.g.][]{prieto08,katz14,li14}. These models reproduce the peaked shape just like the model presented here. However, they retrieve a universal peak mass by construction, because the dominant disruption mechanism in these models is an environmentally-independent formulation of evaporation, in strong contrast with observations, theory, and simulations \citep[e.g.][]{baumgardt03,lamers05,kruijssen11}. Our model does not make this assumption and instead reproduces the GCMF mainly by including the (dominant) evolutionary phase in the natal galaxy disc, while also accounting for the environmental dependence of the evaporation rate in the host galaxy halo. Owing to the important role of the rapid-disruption phase, the model does not predict a radial gradient of the peak mass or require extremely anisotropic GC orbits, both of which were predicted by earlier, evaporation-only models but are unobserved \citep[e.g.][]{fall01,vesperini03}.

The dependence of the specific frequency on galaxy mass or metallicity is rarely considered in GC formation models. A recent example is presented in \citet{katz14}, who find a decreasing number of GCs per unit host stellar mass. However, this is a direct consequence of their assumption of environmentally-independent GC disruption and their GC formation model, which assumes that GCs form in direct proportion to the dark matter halo mass and thus automatically represent a lower fraction of the stellar mass in galaxies with high stellar mass fractions. The specific frequency is therefore not predicted by the model, but follows directly from the assumptions. In model presented here, we do not make these assumptions, but predict the specific frequency using a physical description in which GCs form from the ISM in galaxy discs ({\it not} in proportion to the halo mass) and undergo environmentally-dependent disruption, both of which are consistent with the variety of constraints presented in \S\ref{sec:constr}.

Likewise, previous models reproduce the colour and metallicity distributions of GCs by assuming a relation between the number of GCs and the host galaxy stellar or halo mass, either in combination with an environmentally-independent disruption model \citep[e.g.][]{muratov10,katz14,li14} or without including a disruption model \citep{tonini13}. In these models, the metal-poor GC population originates from the dwarf galaxies that are accreted during hierarchical galaxy growth, which have high specific frequencies and therefore lead to a peak at low metallicities, even though such a peak is not present in the field star metallicity distribution. Our model does not a priori assume any scaling between the number of GCs and the host galaxy properties, and instead provides the physical explanation for these scaling relations.

In view of the model's simplicity, it is particularly compelling that it simultaneously reproduces a substantial number of observables. This is encouraging and provides a fruitful starting point for future work, especially given the many predictions of the model, which are discussed in \S\ref{sec:pred}.

\section{Predictions and implications for globular cluster formation and evolution} \label{sec:pred}
Having tested the two-phase model for GC formation against several of the key characteristics of GC populations in the local Universe, it is appropriate to outline its implications, as well as to provide quantitative predictions for future work.

\subsection{Most of the cluster disruption takes place during the rapid-disruption phase in the host galaxy disc} \label{sec:discdis}
\begin{figure*}
\center\includegraphics[width=\hsize]{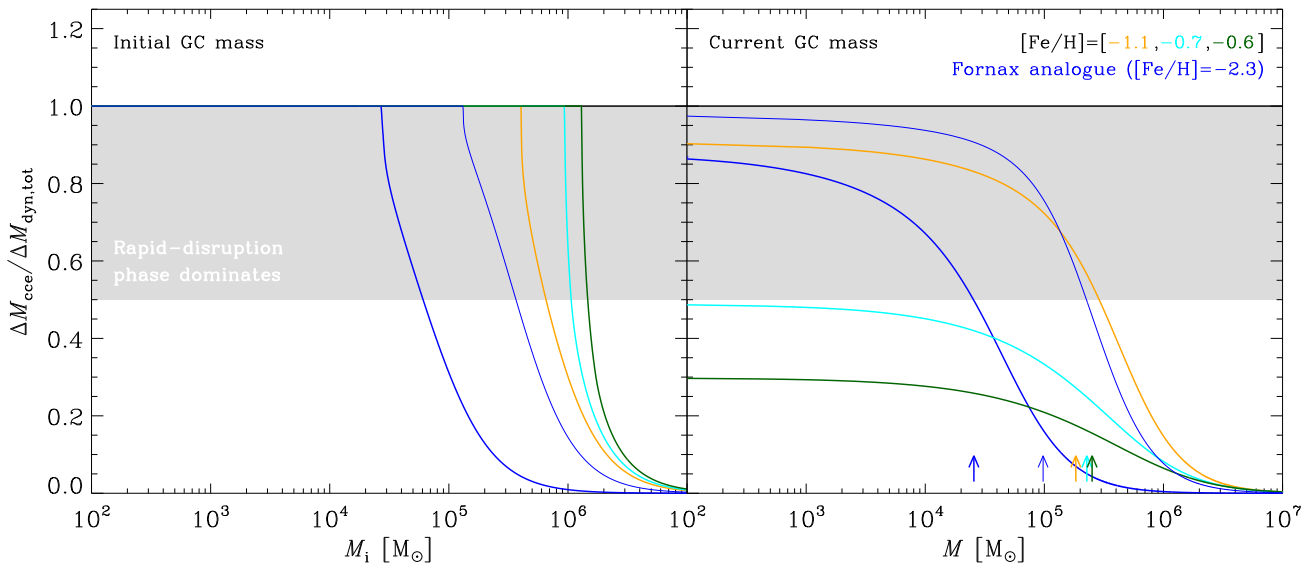}
\caption{This figure shows that the majority of cluster disruption takes place during the rapid-disruption phase in the host galaxy disc. Shown is the ratio between the dynamical mass loss during the rapid-disruption phase and the total dynamical mass loss $\Delta M_{\rm cce}/\Delta M_{\rm dyn,tot}$ as a function of the initial GC mass $M_{\rm i}$ (left panel) and the current GC mass $M$ (right panel) for the five different galaxy models from Table~\ref{tab:init}. The legend indicates the colour coding, where the \{thin,thick\} blue lines represent the \{long,short\}-$t_{\rm merge}$ Fornax model (towards the right and left in the figure, respectively). The grey-shaded area indicates where the dynamical mass loss during the rapid-disruption phase in the disc exceeds that during the slow-disruption phase in the halo. For reference, the upward-pointing arrows in the right-hand panel denote the peak masses $M_{\rm peak}$ (which roughly coincide with the median masses) of the $z=0$ GCMFs from Figure~\ref{fig:gcmf}. Clusters with $\Delta M_{\rm cce}/\Delta M_{\rm dyn,tot}=1$ in the left-hand panel are destroyed during the rapid-disruption phase, which occurs for the vast majority of the initial population. Omitting the short-$t_{\rm merge}$ Fornax model, only clusters with initial masses $M_{\rm i}\ga10^6~\msun$ (corresponding to current masses $M\gg M_{\rm peak}$) or metal-rich, surviving GCs ($\feh\geq-0.7$) experienced most of their dynamical mass loss by evaporation in the host galaxy halo.} \label{fig:phases}
\end{figure*}
As a function of both the initial and current GC masses, Figure~\ref{fig:phases} shows the mass loss due to tidal shocking by GMCs during the rapid-disruption phase in the host galaxy disc relative to the total dynamical mass loss integrated over the total GC lifetime. It is clear that the two-phase GC formation model predicts that most of the cluster disruption occurs in the host galaxy disc, i.e.~during the rapid-disruption phase of Figure~\ref{fig:tschem}. During this phase, clusters reside in the gas-rich component of the host galaxy and are disrupted by the tidal injection of energy during encounters with GMCs. All clusters with initial masses below $M_{\rm i}=3\times10^4~\msun$ are disrupted during this rapid-disruption phase, i.e.~before they escape the host galaxy disc. For the most metal-rich clusters, the minimum mass necessary to survive the rapid-disruption phase may have been as high as $M\sim10^6~\msun$. Only the presently most massive ($M\gg M_{\rm peak}$) or metal-rich ($\feh\geq-0.7$) GCs may have had such short dynamical times that they could respond adiabatically to the tidal perturbations, allowing them to experience most of their mass loss during the slow-disruption, evaporation-driven phase in the galaxy halo.

\subsection{Relations between the number of GCs and their host galaxies are in place at $z=1$--$2$}
Because most of the cluster disruption takes place due to tidal shocking by GMCs in the host galaxy disc, we predict that the well-known relations between the number of GCs and their host galaxy properties (e.g.~stellar mass, halo mass, or metallicity, see Figures~\ref{fig:tnfeh} and~\ref{fig:mgcmh}) were already in place at $z=1$--$2$, whereas the final masses and radii of GCs were attained by their quiescent evolution in galaxy haloes. Likewise, the GCMF at $z=1$--$2$ should already have a peak mass similar to (or even higher than) the characteristic GC mass-scale seen at the present day, i.e.~$M_{\rm peak}\ga10^{5.3}~\msun$ for $z<2$ (see Figure~\ref{fig:gcmf}). This prediction strongly contrasts with models explaining the peaked GCMF by relaxation-driven evaporation, for which the peak mass should decrease by 0.3~dex for every factor of two in cosmic time \citep[e.g.][]{fall01}, yielding peak masses of $M_{\rm peak}=\{10^5,10^{4.8},10^{4.3}\}~\msun$ for $z=\{0.5,1,2\}$, assuming a typical formation redshift of $z=3$. The peak mass required by the evaporation model at $z=2$ is at least an order of magnitude lower than predicted by our model.

By the time GCs have escaped into the host galaxy halo (which is typically after a few~\gyr, see Table~\ref{tab:init}), they will have faded substantially compared to their initial luminosities. With HST, it is therefore not possible to detect the GC peak mass at $z=0.5$--$2$. Even with the James Webb Space Telescope (JWST), a signal-to-noise ratio of ${\rm S}/{\rm N}>1$ may only be reached out to $z=0.5$ for massive ($M>10^5~\msun$) GCs, requiring multi-hour exposures. Similar exposure times will be required for thirty-metre class telescopes such as the European Extremely Large Telescope (E-ELT) and the Thirty Meter Telescope (TMT). None the less, the intermediate-redshift evolution of the GCMF is a critical discriminator between the evaporation and two-phase GC formation models once observational facilities gain the required resolution and sensitivity.

\subsection{GCs were formed according to the same power-law ICMF as young clusters in the local Universe} \label{sec:icmf}
The prediction that the high, present-day GCMF peak mass was already reached at $z=0.5$--$2$ is not unique -- the same prediction is made by models proposing that the ICMF of GCs was peaked at birth \citep[e.g.][]{parmentier07}. However, in our model, it is assumed that the ICMF follows the same power-law mass function as seen for cluster formation in the local Universe. The good agreement of the modelled GC populations with those observed at $z=0$ lends some support to this assumption -- it suggests that GCs indeed formed according to the same power-law ICMF as young clusters in the local Universe. As shown by \citet{katz13}, the YMC progenitors of GCs should be readily observable with JWST in the formation redshift range $z=2.3$--$8$. The measured ICMF should allow us to distinguish between the presented model, in which the ICMF follows a power law, and those models proposing that the GC peak mass is an imprint of their formation process.

\subsection{The high-mass ends of the GMC and cluster mass functions at high redshift are set by the Toomre mass}
The two-phase GC formation model reproduces the properties of observed GC populations by assuming that the maximum mass-scale of the ICMF is related to the maximum mass-scale for gravitational instability (the Toomre mass) in the same way as holds in nearby star-forming galaxies. The associated prediction is that the high-mass end of the GMC mass function in $z>2$ galaxies is set by the Toomre mass. With ALMA, it will be possible to observe the high-mass end of the cloud mass function in these galaxies, which will provide a direct test for the relation between the Toomre mass $M_{\rm T}$ and the GMC mass. In combination with the aforementioned measurement of the ICMF (\S\ref{sec:icmf}), it will also verify the adopted relation between the maximum GMC mass and the maximum YMC mass $M_{\rm c}$, which is assumed to satisfy $M_{\rm c}\equiv M_{\rm cl,T}=\epsilon\Gamma M_{\rm T}$ (see \S\ref{sec:init}).

\subsection{The cosmic GC formation history peaks at a higher redshift than the cosmic SFH}
Galaxy mergers play a central role in our model, because they enable the migration of clusters out of the host galaxy disc into the galaxy halo. As has been discussed at length, this process strongly increases the long-term survival chances of the clusters. Because the merger rate increases with redshift, the cosmic GC formation history should peak at a higher redshift than the cosmic star formation history. Earlier estimates by  \citet[Figure~20]{peng08} suggest that the redshift difference is roughly $\Delta z\sim1$ (i.e.~$\Delta t\sim1~\gyr$ for $z=2$--$4$) and thus should be detectable given similarly-accurate GC age estimates.

\subsection{There is a time delay between cluster formation and their migration into the galaxy halo} \label{sec:delay}
Our model assumes a delay of several $\gyr$ between cluster formation in galaxy discs and their migration by mergers and/or stripping into the halo, which plays a critical role in determining the properties of the GC population at $z=0$. The good agreement between this simple model and the observations suggests that GC formation mechanisms that deposit GCs near-instantly in the halo are sub-dominant. As a result, either (1) the clusters that are formed during a merger itself have a low probability of escaping the host galaxy because they are destroyed on time-scales shorter than a dynamical time (see \S\ref{sec:phys}), or (2) the majority of all GC migration occurs through the tidal stripping of GCs from their host galaxies by more massive haloes, in which case migration is rarely associated with a starburst event. 

The first of the above points would suggest that all clusters that formed at least one galactic dynamical time before a major merger can safely migrate into the halo. Observations of the galaxy merger pair NGC~4038/39 (the Antennae galaxies) by \citet{bastian09} show that surviving clusters without disc kinematics (indicative of migration) indeed have ages of $200$--$500~\myr$, which is confirmed by numerical simulations of galaxy mergers \citep{kruijssen11c}. This indicates a formation time that precedes the most recent encounter by $>100~\myr$ and thus supports point (1). The second point can be tested using the mass ratio-dependence of the merger rate in the Millennium Simulation. \citet{genel09} show that at $z\sim3$, only galaxies with halo masses $M_{\rm h}<10^{12}~\msun$ typically merge with more massive galaxies and thus lose their GCs by stripping. At higher halo masses (corresponding to $\feh>-0.8$, see Table~\ref{tab:init}), the majority of the merging takes place with haloes of similar masses, indicating that the migration of GCs by the tidal stripping of their host galaxies is not significant. This metallicity cut range covers a minority of the GC metallicity range, confirming that point (2) above typically holds too.

\subsection{The increase of the specific frequency at high galaxy masses is caused by a combination of satellite galaxy accretion and merger-induced cluster formation}
We see that GCs with metallicities $\feh>-0.8$ have migrated into the halo by major mergers, which is possibly associated with a starburst if the host galaxy is gas-rich, whereas lower-metallicity GCs have migrated by tidal stripping. In the Virgo Cluster, the mean metallicities of the GC populations reach $\feh\sim-0.8$ for a host galaxy magnitude $M_B\sim-20$ (or a stellar mass $M_{\rm star}\sim10^{11}~\msun$), coinciding exactly with the magnitude at which the specific frequency $T_N$ reaches a minimum (see Figure~6 of \citealt{peng08}). As discussed in \S\ref{sec:tn2}, it is likely that the increase of $T_N$ in more massive galaxies ($M_{\rm star}>10^{11}~\msun$) at least partially arises due to the accretion of GCs from lower-mass galaxies, but an additional contribution may come the change of GC migration agent discussed here. If the migration of clusters in haloes with masses $M_{\rm h}>10^{12}~\msun$ is associated with a starburst in which large numbers of clusters are born, at least some of the newly-formed clusters could find safety in numbers and survive despite the low survival probability of clusters born at the peak SFR of galaxy mergers.

Future work can quantify the relative contributions of both mechanisms for increasing $T_N$ at high galaxy masses with a number of simple measurements. Firstly, the mass fraction accreted in the form of dwarf galaxies (which have high $T_N$) should increase with galaxy mass more steeply than $M_{\rm acc,tot}\propto M_{\rm star}^{0.3}$ to offset the intrinsic decrease of $T_N$ with galaxy mass (see Figure~\ref{fig:tnfeh}). Secondly, the metallicities of the GCs leading to the increase of the specific frequency at high galaxy masses differs strongly for both mechanisms. If tidal stripping dominates, the extra GCs should predominantly be metal-poor and the increase of the mean GC metallicity with galaxy mass should diminish. If the migration event is associated with a starburst that supplies additional GCs, the extra GCs should be metal-rich and the correlation between mean GC metallicity and galaxy mass should steepen. Figure~9 of \citet{peng08} shows that $T_N$ increases both for the metal-rich and the metal-poor GCs for galaxies with $M_B<-20$, suggesting that both mechanisms should be at work.

\subsection{The specific frequency increases with galaxy clustering}
The galaxy merger time-scale varies with the cosmic environment, e.g.~between highly clustered and (near-)isolated field galaxies. A general prediction of the presented model is that the specific frequencies of field galaxies (which have low merger rates) should be systematically lower than those of galaxies in dense environments (which have high merger rates). This environmental dependence of $T_N$ arises because the modelled duration of the rapid-disruption phase is set by the galaxy merger time-scale. In other words, the specific frequency of otherwise identical progenitor galaxies with different merger time-scales should scale as $T_N\propto t_{\rm merge}^{-1}$. Observationally, this variation can be constrained by considering GCs at fixed metallicity, which removes the dependence of $T_N$ on the galaxy mass at GC formation seen in Figure~\ref{fig:tnfeh}. Indeed, \citet{peng08} find an increase of the specific frequency towards the centre of the Virgo Cluster at fixed $g-z$ colour (i.e.~at fixed GC metallicity), which at least qualitatively confirms the dependence of the GC survival fraction on the galaxy merger rate that is predicted by the proposed, two-phase model. Note that even if galaxy mergers are not responsible for the migration of GCs into a gas-poor environment, but instead the gas is removed (e.g.~by feedback, strangulation, or ram pressure stripping, \citealt{vandenbosch08}), then the enhanced quenching of star formation activity observed in galaxy clusters (`environment quenching', see e.g.~\citealt{peng10}) suggests that the rapid-disruption phase may have a shorter duration in clustered environments. (although the time-scales may be of a similar order of magnitude, for e.g.~strangulation see \citealt{peng15}). In addition, such early quenching of star formation prevents the further growth of the field star population after the majority of the GCs has formed, which further boosts the specific frequency. All of these mechanisms should lead to a trend of an increasing specific frequency with galaxy clustering.

\subsection{The GC evaporation time-scale decreases with the GC metallicity} \label{sec:gradient}
To obtain a near-universal GCMF, our model requires a weak negative metallicity gradient of a galaxy's GC population, such that the disruption time-scale due to evaporation decreases with the metallicity (see Table~\ref{tab:init} and Figure~\ref{fig:gcmf}). In \S\ref{sec:gcmf2}, we showed that the required gradient is quantitatively consistent with that observed for GCs in the Milky Way. It was proposed that this gradient is an imprint of the galaxy mass (and hence metallicity) dependent redistribution of GCs in mergers and/or tidal stripping, in which metal-rich GCs formed at higher binding energies (and experience stronger tidal fields). If true, then the relation between $\feh$ and the disruption time-scale due to evaporation provided in equations~(\ref{eq:fehobs}) and~(\ref{eq:fehmodel}) may hold universally in galaxies where dynamical friction on GCs is negligible (also see \S\ref{sec:df}).

\subsection{Fossil stellar streams in the halo from evaporated GCs are more metal-rich than the surviving GCs}
Observational missions and surveys such as Gaia, SEGUE and RAVE are capable of identifying and characterising the remnants of tidally disrupted stellar systems in the Galactic halo, be it tidal debris from disrupted dwarf galaxies or from evaporated GCs. It was shown in \S\ref{sec:discdis} and the right-hand panel of Figure~\ref{fig:phases} that for metallicities $\feh<-0.7$ most of the mass loss takes place during the rapid-disruption phase in the host galaxy disc, implying that little coherent debris of these clusters should be preserved in the Galactic halo. By contrast, metal-rich GCs ($\feh\geq-0.7$) should typically have lost most of their mass by evaporation in the halo. This is a corollary of \S\ref{sec:gradient}, which states that the GC evaporation time-scale decreases with metallicity. As we will discuss in \S\ref{sec:mi}, GCs with metallicities $\feh\geq-0.7$ and typical masses ($M\sim10^5~\msun$) may have lost $90\%$ of their initial mass, in strong contrast with the substantially lower mass loss ($\la50\%$) of lower-metallicity GCs at the same present-day mass. As a result, the tidal debris of disrupted GCs in galaxy haloes should be dominated by metallicities larger than the median GC metallicity. With Gaia and its spectroscopic companion surveys (e.g.~Gaia-ESO), it will be possible to systematically test this prediction for the Galactic halo.

\subsection{The radii of GCs are set by structural evolution and are unrelated to GC formation or disruption}
While our model predicts that most of the cluster {\it disruption} occurs during the disc phase of Figure~\ref{fig:tschem}, most of the GCs' {\it structural evolution} takes place during the halo phase. As a result, the empirical relation seen in present-day GC populations between the GC half-mass density and GC (peak) mass \citep[$M\propto\rho_{\rm h}^{1/2}$, see e.g.][]{fall77,gnedin97,mclaughlin08} is caused by the evolution of (primarily) the GC radius (and to a limited degree the GC mass) in the host galaxy halo. The variation of the GC mass with the half-mass density is therefore unrelated to the conditions at GC formation. This is to be expected theoretically -- unless initially $r_{\rm h}\propto M^\beta$ with $\beta<-1/3$ (at odds with theory and observations, see \S\ref{sec:ymc}), the half-mass relaxation time increases with the cluster mass, causing low-mass clusters to expand more rapidly than massive clusters. After a time $t_{\rm halo}$ of free expansion in the host galaxy halo, this leads to a relation between the half-mass density and the cluster mass of $M\propto\rho_{\rm h}^{1/2}$ \citep{gieles11b}, which is roughly consistent with the increase of the peak mass with the half-mass density from \citet{mclaughlin08}. The observation that the typical GC mass increases with the half-mass density is thus not causal in the sense that high-density GCs dissolve more rapidly, but simply arises due to their structural evolution \citep{gieles11b}.

\subsection{Dynamical friction has affected only a minority of the GC population} \label{sec:df}
Given the properties of the host galaxy, a minimum cluster mass-scale can be derived above which the dynamical friction time-scale to the centre of the galaxy is shorter than the host galaxy merger time-scale, in which case the clusters are destroyed by dynamical friction before they can escape into the halo. Using equation~(6) of \citet{kruijssen14c}, we obtain
\begin{equation}
\label{eq:mdf}
M_{\rm df,min}=10^6~\msun~\left(\frac{t_{\rm merge}}{2~\gyr}\right)^{-1}\left(\frac{R}{2~\kpc}\right)^2\left(\frac{V}{200~\kms}\right) .
\end{equation}
Setting the circular velocity equal to the virial velocity of the halo at $z=3$, we have $V=[10GH(z)M_{\rm h}]^{1/3}$, where $H(z)$ is the Hubble constant at redshift $z$, i.e.~$H(z)=H_0[\Omega_{\rm m}(1+z)^3+(1-\Omega_{\rm m}-\Omega_\Lambda)(1+z)^2+\Omega_\Lambda]^{1/2}$, which in combination with the angular velocity $\Omega$ as a function of stellar (and hence halo) mass from equation~(\ref{eq:omega}) also defines the radius $R=V/\Omega$ as a function of galaxy mass. Substitution into equation~(\ref{eq:mdf}) then yields minimum mass-scales for dynamical friction into the galaxy centre\footnote{To calculate these numbers, we use the Planck 2013 cosmological parameters, with $H_0=67.3~\kms~{\rm Mpc}^{-1}$, $\Omega_{\rm m}=0.315$ and $\Omega_\Lambda=0.685$ \citep{planck14}.} ranging from $10^{5.8}~\msun$ for the late-merger Fornax model to $10^{6.5}$--$10^{8.3}~\msun$ for all other galaxy models, indicating that dynamical friction may have affected a small part of the GC population (thereby possibly fueling the formation of nuclear clusters), but never dominated during early GC evolution. This confirms the result of \citet{jordan07}, who found that dynamical friction cannot explain the exponential truncation mass at the high-mass end of the GCMF.

\subsection{Typical GCs were only 2--3 times more massive at birth} \label{sec:mi}
Considering that the two-phase GC formation model reproduces several of the observed properties of GC populations, it may be used to constrain GC formation theories aiming to reproduce the chemical abundance patterns seen in Galactic GCs. In \S\ref{sec:gc}, we very briefly summarised some of the main assumptions, predictions, and problems of these models. One of these is the {\it mass budget problem}, requiring GCs to have been much more massive at birth than they are today, i.e.~by a factor of $10$--$100$ (for explaining the Na-O anti-correlation in the AGB enrichment model) or by a factor of $>20$ (for forming a sufficiently massive enriched stellar population in the FRMS enrichment model). The two-phase GC formation model allows us to predict initial-to-current mass ratios $M_{\rm i}/M$ for GCs across the observed GC mass and metallicity range and test the assumptions of these enrichment models.

\begin{figure}
\center\includegraphics[width=\hsize]{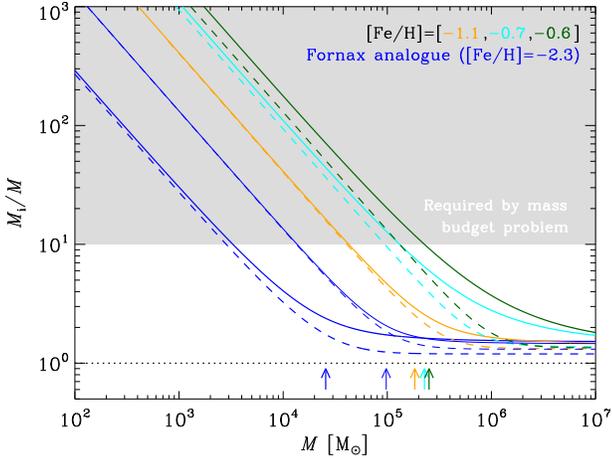}
\caption{This figure shows that only a minority of the present-day GC population satisfies the requirement of the AGB and FRMS enrichment models that GCs were $>10$ times more massive at birth. Shown is the initial-to-current GC mass ratio $M_{\rm i}/M$ as a function of the current GC mass $M$ for the five different galaxies from Table~\ref{tab:init}. The legend indicates the colour coding, where the \{thin,thick\} blue lines represent the \{long,short\}-$t_{\rm merge}$ Fornax model (towards the top right and bottom left in the figure, respectively). Dashed lines show the mass ratio at the end of the rapid-disruption phase ($t_{\rm merge}$) and solid lines show the mass ratio at $z=0$ ($t_{\rm now}$). The horizontal dotted line represents zero mass loss ($M_{\rm i}/M=1$) and the grey-shaded area indicates where the initial mass is $>10$ times larger than the current GC mass. For comparison to the solid lines, the upward-pointing arrows denote the peak masses $M_{\rm peak}$ (which roughly coincide with the median masses) of the $z=0$ GCMFs from Figure~\ref{fig:gcmf}. The condition that GCs were $>10$ times more massive at birth is only satisfied up to a maximum mass $M_{\rm \times10,max}$, which increases from $M_{\rm \times10,max}\sim10^{3.5}~\msun$ for the short-$t_{\rm merge}$ Fornax model to $M_{\rm \times10,max}\sim10^{5.3}~\msun$ for the $\feh=-0.6$ model. Note that $M_{\rm \times10,max}<M_{\rm peak}$ for most of the metallicity range ($\feh<-0.6$), showing that the majority of GCs had initial masses only a few times larger than their present-day masses.} \label{fig:mi}
\end{figure}
Figure~\ref{fig:mi} shows $M_{\rm i}/M$ as a function of the present-day GC mass $M$ for the five galaxy models of Table~\ref{tab:init} at the time of GC migration and at the present day ($t_{\rm merge}$ and $t_{\rm now}$ as defined in Figure~\ref{fig:tschem}). We see that the initial-to-current mass ratio is a function of the current GC properties, decreasing strongly with the GC mass and increasing with the metallicity. There is little evolution between $t_{\rm merge}$ and $t_{\rm now}$, which again indicates that most of the disruption took place in the GCs' high-redshift birth environment. Only low-mass and/or metal-rich GCs may have lost $>90\%$ of their initial mass. If we assume the simple case that both the AGB and FRMS enrichment scenarios require GCs to have been at least 10 times more massive at birth (which is a strict lower limit for both models, see \S\ref{sec:gc}), then Figure~\ref{fig:mi} provides an upper limit to the present-day GC mass $M_{\rm \times10,max}$ below which the AGB or FRMS enrichment models may have operated. This mass limit is typically much lower than the peak mass of the GCMF. Only at metallicities $\feh\geq-0.7$ we find $M_{\rm \times10,max}\sim M_{\rm peak}$.

At metallicities $\feh<-0.7$ the majority of the GC population has a mass loss history that is inconsistent with the AGB or FRMS enrichment models, as most GCs were much less than 10 times more massive at birth. As explained in \S\ref{sec:gc}, both enrichment models assume that most of the unenriched stars are lost in order to avoid the mass budget problem. In other words, they predict that the enriched fraction must increase with the initial-to-final mass ratio, because it reflects the amount of mass loss. Combining this prediction with the results of Figure~\ref{fig:mi}, we see that according to the AGB and FRMS models the enrichment fraction should decrease with the GC mass and increase with the metallicity, becoming very small for $M>10^5~\msun$ and $\feh<-0.7$. The contrary is observed -- as discussed in \S\ref{sec:gc}, chemically enriched stellar populations have been observed in all considered GCs with masses $10^{5}\la M/\msun\la 10^{6.5}$ and metallicities $-2.3<\feh<-0.3$, and the enriched fraction {\it increases} with the GC mass \citep[see e.g.~Figure~2 of][]{conroy12}.

It was shown in \S\ref{sec:comp} that the predicted initial-to-current mass ratios of Figure~\ref{fig:mi} are consistent with the observed ratios between the number of GCs and the field star population (i.e.~the specific frequency). If the GCs would have attained their present-day masses by losing more mass, then a larger field star population would have been necessary. In other words, the host galaxies simply do not contain enough stellar mass for the GCs to have lost more mass than predicted by the two-phase model, even though at first sight the small fraction of the total stellar halo mass constituted by GCs may suggest otherwise (see \S\ref{sec:mmin}). As we have seen in \S\ref{sec:constr} and \S\ref{sec:model}, the reason for the low GC mass fraction in galaxy haloes is that during a star formation event, only some fraction of the stellar mass ends up in the clusters with masses $M\geq10^6~\msun$ that could survive to become GCs. As a result, galaxy haloes throughout the Universe would need to contain several times more field stars than observed to be consistent with the AGB and FRMS enrichment models, unless the stellar initial mass function, the CFE or the ICMF were altered, in conflict with theory and observations. It is therefore more likely that the fraction of enriched stars reflects some aspect of the conditions at GC formation (as it does in the EDA enrichment model) than that these stars formed from AGB or FRMS ejecta in a second star formation episode.

\subsection{The minimum mass for the formation of multiple stellar populations in GCs was $\sim10^5~\msun$ at $z>2$} \label{sec:mmin}
All GC enrichment models discussed in \S\ref{sec:gc} suggest that the presence of a chemically enriched stellar population requires some minimum cluster mass (or density, which for $r_{\rm h}\propto M^\beta$ with $\beta<1/3$ can still be translated to a minimum mass). By comparing the total enriched mass in the Galactic stellar halo to that in GCs, it is possible to empirically constrain this minimum mass for cluster enrichment. About 2\% of the field star population in the Galactic halo consists of enriched stars \citep[e.g.][also see \citealt{vesperini10}]{carretta10b}. This corresponds to a mass of $\sim2\times10^7~\msun$ \citep[e.g.][]{bell08}, similar to the total enriched stellar mass presently in GCs (also $1$--$2\times10^7~\msun$, assuming that presently 50\% of the GC stars are enriched). As such, a total of $\sim4\%$ of all Galactic halo stars are enriched. If we assume that enrichment only took place in clusters above some minimum initial mass, we can use the mass loss histories provided by our two-phase GC formation model to determine the minimum mass for GC enrichment to produce a stellar halo in which a total of $4\%$ of all stars are enriched.\footnote{The answer to this question necessarily relies on the relative escape rates of stars from the enriched and non-enriched populations. If non-enriched and enriched stars have been lost from GCs at a similar rate throughout cosmic history, then the similar total enriched stellar masses in the halo and in GCs suggests that enriched stars formed in GCs above some minimum mass such that {\it on average} they lost half of their mass through dynamical mass loss. In principle, this minimum mass could be estimated for each of the model galaxies considered here. However, the three GC enrichment models also predict that the enriched population should be more centrally concentrated than the non-enriched population. This chemical segregation causes non-enriched stars to be lost preferentially, which may increase the ratio of enriched to non-enriched stars by up to a factor of two \citep[e.g.][]{decressin08}.}

To address the origin of the enriched halo stellar population, we first determine what fraction $f_{\rm GC,i}$ of the entire coeval stellar population is constituted by stars that were born in (but possibly lost from) the present, surviving GCs. For the five galaxy models considered here, this number varies in the range $f_{\rm GC,i}=2$--$16\%$, with a median of $f_{\rm GC,i}=7\%$ -- the other 93\% was never part of the GCs that survived to the present day. Given that the present GC enrichment ratio (${\rm enriched}:{\rm unenriched}\sim1:1$) is representative of the initial enrichment fraction in GCs to within a factor of 2 \citep{decressin08}, a fraction $1/3$--$2/3$ of the lost stars may be enriched, i.e. $1/3$--$2/3\times7\%\sim2$--$5\%$ of the stellar halo consists of enriched stars that once resided in surviving GCs. This matches the observed enriched fraction in the Galactic halo and therefore requires that all of the enriched stars have been supplied by the present GCs. The minimum initial mass of the surviving GCs $M_{\rm i,min}$ should thus reflect the minimum cluster mass required for enrichment $M_{\rm enr,min}$.

Our five galaxy models yield $M_{\rm i,min}\equiv M_{\rm enr,min}=10^{4.5}$--$10^{6.2}~\msun$ (see Figure~\ref{fig:phases}) with a median of $M_{\rm enr,min}=10^{5.5}~\msun$.\footnote{If enrichment occurs in clusters above some critical half-mass density rather than above some critical mass, it is possible that the corresponding mass limit is higher in the local Universe than it was at the time of GC formation. The initial radii of YMCs may increase with decreasing pressure or ambient density (see \S\ref{sec:constr}), and hence with cosmic time. As explained in \S\ref{sec:model}, we have scaled the YMC radius by the mean tidal density, leading us to adopt radii at $z=3$ a factor of $2$--$3$ smaller than those seen at $z=0$. In that specific case, the present-day threshold mass for the formation of an enriched population would currently be $\sim10$ times larger than at $z=3$, i.e.~$M_{\rm enr,min}\ga10^6~\msun$.} This high mass is required by the small fraction of enriched stars in the Galactic halo, and is uncertain by a factor of $\sim2$ depending on how representative the present GC enrichment fraction is of the initial GC population \citep{decressin08}. Considering that most of the surviving GCs had initial masses exceeding this limit, it is possible that multiple stellar populations are indeed a universal property of GCs.

\subsection{Most GCs could not have survived the high gas densities required for pristine gas accretion} \label{sec:acc}
Finally, we turn to the {\it gas accretion problem} of the AGB enrichment model, which requires a pristine gas mass comparable to the initial GC mass to be accreted for the formation of the enriched stellar population (see \S\ref{sec:gc}). A possible mechanism for supplying the gas \citep[e.g.][]{conroy11,maxwell14} is Bondi-Hoyle-Lyttleton (BHL) accretion \citep{hoyle41,bondi44,bondi52}, which occurs at a rate
\begin{equation}
\label{eq:bhl}
\left(\frac{{\rm d}M}{{\rm d}t}\right)_{\rm BHL}=\frac{2\alpha G^2M^2\rho_{\rm ISM}}{(v_{\rm rel}^2+c_{\rm s}^2)^{3/2}} ,
\end{equation}
where $\alpha$ is a constant of order unity, $v_{\rm rel}$ is the relative velocity between the accreting body and the surrounding material, and $c_{\rm s}$ is the sound speed. One can define a time-scale for accreting the initial mass of the accretor by writing $t_{\rm BHL}\equiv M/({\rm d}M/{\rm d}t)_{\rm BHL}\propto v_{\rm rel}^{3}M^{-1}\rho_{\rm ISM}^{-1}$ (where we have assumed $v_{\rm rel}\gg c_{\rm s}$). This expression shows that accretion proceeds more rapidly in a higher-density ISM. However, the disruption time-scale due to tidal shocks has a very similar scaling of $t_{\rm cce}\propto M\rho_{\rm ISM}^{-3/2}$, indicating that a higher-density ISM disrupts clusters more efficiently. Because the density-dependence of the disruption time-scale is steeper than that of the accretion time-scale, there must exist a maximum ISM density below which significant gas accretion can occur before a cluster of mass $M$ is disrupted, i.e.~$t_{\rm BHL}/t_{\rm cce}\propto M^{-2}\rho_{\rm ISM}^{1/2}<1$. Vice versa, given an ISM density there exists a minimum mass above which a cluster can survive long enough to accrete.

\begin{figure}
\center\includegraphics[width=\hsize]{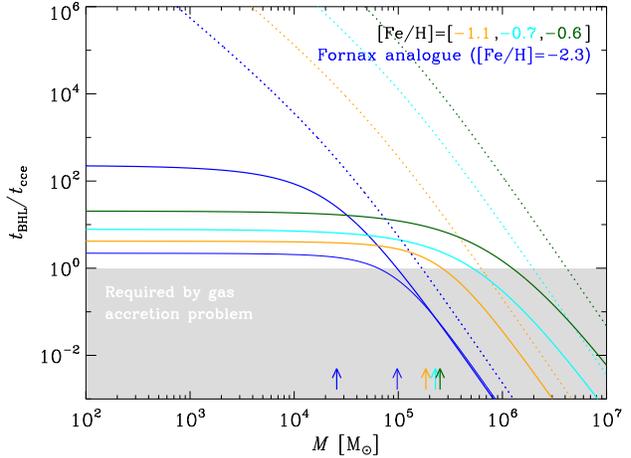}
\caption{This figure shows that only a minority of the present-day GC population can satisfy the requirement of the AGB enrichment model that GCs should accrete a pristine gas mass similar to their initial mass. Shown is the time-scale ratio between the Bondi-Hoyle-Lyttleton (BHL) accretion time-scale and the disruption time-scale during the disc phase (see Figure~\ref{fig:tschem}), as a function of the initial (dotted lines) and current (solid lines) GC masses for the five different galaxies from Table~\ref{tab:init}. The legend indicates the colour coding, where the \{thin,thick\} blue lines represent the \{long,short\}-$t_{\rm merge}$ Fornax model (towards the bottom and top in the figure, respectively). The grey-shaded area indicates where $t_{\rm BHL}<t_{\rm cce}$, which is required for gas accretion to outpace cluster disruption by the high-density environment. For comparison to the solid lines, the upward-pointing arrows denote the peak masses $M_{\rm peak}$ (which roughly coincide with the median masses) of the $z=0$ GCMFs from Figure~\ref{fig:gcmf}. The condition for pristine gas accretion is only satisfied above some minimum mass $M_{\rm acc,min}$ that exceeds the GCMF peak mass ($M_{\rm acc,min}>M_{\rm peak}$) for almost all metallicities.} \label{fig:fcceacc}
\end{figure}
The competition between accretion and disruption is visualised in Figure~\ref{fig:fcceacc}, which shows the time-scale ratio $t_{\rm BHL}/t_{\rm cce}$ as a function of the initial (dotted) and present-day (solid) GC mass for the five galaxy models of Table~\ref{tab:init}. For the relative velocity we have adopted the circular velocity $V$ (rather than a velocity dispersion), which is appropriate for clusters on plunging, eccentric orbits \citep[as proposed by][]{maxwell14}. Figure~\ref{fig:fcceacc} shows that the conditions necessary for significant BHL accretion are so disruptive that only clusters with very high initial masses may have survived the process. Pristine gas accretion without disrupting the cluster is not possible below a mass-scale $M_{\rm acc,min}$, which increases with metallicity from \{initial,~current\} masses of $M_{\rm acc,min}\sim\{10^{5.2},10^{4.8}\}~\msun$ at $\feh=-2.3$ to $M_{\rm acc,min}\sim\{10^{6.6},10^{6.1}\}~\msun$ at $\feh=-0.6$. Because this analysis strictly concerns the dynamics of gas accretion and tidal disruption while excluding any form of feedback, these masses represent firm lower limits for pristine gas accretion.

Comparing $M_{\rm acc,min}(t_{\rm now})$ to $M_{\rm peak}$ across the galaxy mass range shows that $M_{\rm acc,min}>M_{\rm peak}$ for all galaxies other than the long-$t_{\rm merge}$ Fornax model, indicating that a minority of young GCs can have accreted pristine gas. Moreover, $M_{\rm acc,min}>M_{\rm c}$ in all galaxies other than the Fornax models, suggesting that few clusters massive enough to undergo significant pristine gas accretion ever formed at all. We conclude that high-pressure environments are generally unable to drive the accretion of pristine gas onto young GCs without disrupting them.

At the extremely high GC masses where BHL accretion could possibly proceed without disrupting the cluster, the results of Figure~\ref{fig:fcceacc} imply that the amount of dilution by pristine gas in the AGB enrichment model should increase with the GC mass. In the AGB enrichment model, it is this dilution which turns the Na-O correlation predicted by stellar evolution models into the observed anti-correlation. As we have seen, metal-rich galaxies at $z>2$ are more disruptive to their young GCs than metal-poor ones, hence Figure~\ref{fig:fcceacc} shows that the Na-O anti-correlation should weaken towards increasing metallicity, because the minimum mass for dilution should increase. This suggests that metal-rich GCs like NGC~104, 6388, and 6441 should only have been able to develop a weak Na-O anti-correlation (or none at all) if the AGB enrichment model holds. However, no such dependence on metallicity is observed \citep[e.g.][]{gratton12}.

Finally, comparing $M_{\rm acc,min}$ to the maximum mass $M_{\rm \times10,max}$ below which GCs were $>10$ times more massive at birth (see Figure~\ref{fig:mi}), it is clear that $M_{\rm acc,min}/M_{\rm \times10,max}\ga4$ for all galaxies, indicating that the two conditions required by the AGB enrichment model (allowing substantial mass loss as well as pristine gas accretion) cannot be satisfied simultaneously. This presents a serious challenge to AGB enrichment as the explanation for the observed abundance anomalies.
\begin{table*}
 \centering
  \begin{minipage}{135mm}
  \caption{Predicted GC mass limits for surviving GCs, dynamical friction, gas accretion, and $>90\%$ mass loss.}\label{tab:pred}
  \begin{tabular}{@{}l c c c c c@{}}
  \hline
     & Fornax (1) & Fornax (2) & $\feh=-1.1$ & $\feh=-0.7$ & $\feh=-0.6$ \\
  \hline
  $M_{\rm i,min}~[10^5~\msun]$ & $1.32$ & $0.29$ & $4.08$ & $10.73$ & $17.24$ \\
  $M_{\rm df,min}(t_{\rm form})~[10^5~\msun]$ & $6.82$ & $94.88$ & $44.24$ & $332.79$ & $1938.50$ \\
  $M_{\rm acc,min}(t_{\rm form})~[10^5~\msun]$ & $1.68$ & $1.68$ & $6.24$ & $19.60$ & $41.44$ \\
  $M_{\rm acc,min}(t_{\rm now})~[10^5~\msun]$ & $0.64$ & $0.97$ & $2.63$ & $5.15$ & $12.42$ \\
  $M_{\rm \times10,max}(t_{\rm now})~[10^5~\msun]$ & $0.14$ & $0.03$ & $0.43$ & $1.31$ & $2.27$ \\
  \hline
\end{tabular}
\end{minipage}
\end{table*}

\subsection{Summary of predicted critical GC mass scales}
Throughout this section, we have discussed several predictions of the two-phase GC formation model for future observational tests, as well as its implications for existing GC formation and evolution models. Among these were several minimum and maximum mass-scales above or below which several physical mechanisms could operate. For reference, these mass limits are summarised in Table~\ref{tab:pred}, which shows the minimum initial GC mass of the present-day GCs $M_{\rm i,min}$, the minimum GC mass for dynamical friction to the galactic centre during the disc phase $M_{\rm df,min}$, the minimum initial ($t_{\rm form}$) and current ($t_{\rm now}$) GC masses needed for the accretion of pristine gas during the disc phase $M_{\rm acc,min}$ (as required by the AGB enrichment model), and the maximum current GC mass below which the initial mass was $>10$ times higher at birth $M_{\rm \times10,max}$ (as required by the AGB and FRMS enrichment models). Note that for all models $M_{\rm acc,min}>M_{\rm \times10,max}$, indicating that no GC can both have accreted a substantial amount of pristine gas {\it and} have been $>10$ times more massive at birth.

\section{Open questions and future directions of research} \label{sec:open}
In \S\ref{sec:model}, we used the observational constraints on GC formation from \S\ref{sec:obs} and the theoretical state of the field from \S\ref{sec:phys} to formulate a two-phase model for GC formation. While this model reproduces the main observed properties of GC populations and makes several predictions for future work, its simplicity should not be underestimated. The model describes the big picture of GC formation and evolution until the present day, but there are several open questions that still require answering.
\begin{enumerate}
\item
Based on the pronounced relation between the GMC mass function and the ICMF of star-forming galaxies in the local Universe (see \S\ref{sec:ymc}), we have connected the maximum cluster mass to the Toomre mass, i.e.~the maximum mass-scale below which self-gravity can overcome galactic shear. While this is a reasonable step to make and has been tested in the nearby Universe \citep{kruijssen14c,adamo15b}, observational confirmation is still lacking at high redshift. Are the Toomre mass and the maximum cluster mass related in the same way at $z>2$ as they are at $z=0$? If so, what are the masses of the most massive GCs that ever formed? The most massive clumps seen in $z>2$ galaxies have masses of $\sim10^9~\msun$, consistent with the Toomre mass in these systems \citep[e.g.][]{genzel11,swinbank12}. While these clumps may represent the extreme end of the mass-scale and may not be the most abundant, the obvious question is whether they ever formed any YMCs with masses $M>10^7~\msun$ (i.e.~more massive than the initial masses of most GCs), and whether any such YMCs then escaped the gas-rich component of their host galaxy. Above some mass-scale, the clumps and YMCs may have spiralled in to the galaxy's centre by dynamical friction, but we have seen in \S\ref{sec:pred} that the required minimum mass is well in excess of the exponential truncation mass of the ICMF. None the less, could dynamical friction have moved extreme YMCs into different environments? In the coming years, these questions should be addressable at $z>2$ using ALMA to study massive gas clumps and large-scale telescopes equipped with adaptive optics (e.g.~E-ELT, TMT) to study YMCs.
\item
A key ingredient for reproducing the observed properties of GC populations in nearby galaxies is an increasing ambient ISM density $\rho_{\rm ISM}$ (or more fundamentally its pressure) with host galaxy mass $M_{\rm star}$ during the formation of the GC progenitor clusters. This is required for obtaining the observed dependence of the specific frequency on the metallicity and galaxy stellar mass at formation -- this relation between $\rho_{\rm ISM}$ and $M_{\rm star}$ sets the disruption time-scale during the rapid-disruption phase $t_{\rm cce}$ and thus determines what fraction of the initial cluster population survives until they migrate into the halo by a galaxy merger and/or tidal stripping. How universal is the $\rho_{\rm ISM}$--$M_{\rm star}$ relation? Again, using ALMA to map the molecular gas properties of $z>2$ galaxies will be instrumental in answering this question.
\item
In the presented model, galaxy mergers and the tidal stripping of accreted satellites facilitate the migration of GC-like YMCs from the gas-rich host galaxy disc into the galaxy halo. In the context of hierarchical galaxy formation, it seems indisputable that this process takes place. However, it is very well possible that additional migration mechanisms exist. For instance, secular evolutionary processes such as dynamical interactions with the gas clumps seen in high-redshift galaxies could eject massive clusters from the disc. If the Galactic disc was in place before the formation of the bulge \citep[e.g.][]{bournaud07}, this scenario could explain why the metal-rich GCs in the Milky Way exhibit a net co-rotation with the disc \citep{frenk80}. If multiple processes contribute to the `liberation' of GCs into their host galaxy haloes, the key question is what the relative contributions and characteristic time-scales of these processes are. In the absence of clumpy galaxies in the local Universe to provide any observational constraints, the most promising avenue to address this question will be to quantify the migration mechanisms in high-resolution galaxy formation simulations with sufficient resolution to resolve the formation of gravitationally bound, massive clusters.
\item
As discussed in \S\ref{sec:phys}, it is well-understood how GC radii evolve in the quiescent environment of galaxy haloes. However, it is unclear how the structural evolution of YMCs proceeds in the erratic tidal fields of the high-pressure galaxy discs in which they formed. During the rapid-disruption phase of the two-phase model (see Figure~\ref{fig:tschem}), the initial cluster radii and their time-evolution influence how susceptible the clusters are to disruption by tidal shocks (for which the disruption time-scale increases with the cluster's volume density). At the same time, tidal shocks themselves may induce cluster compression or expansion, suggesting that cluster disruption could be regulated (if tidal shocks cause compression) or could be a runaway process (if tidal shocks cause expansion). In either case, cluster survival is intimately linked not only to their mass evolution, but also to their radius evolution. A related question is whether there exists a maximum stellar volume density above which two-body relaxation acts as a spring that drives expansion and stops further compression. If so, this could lead to a dependence of the cluster radius on the cluster mass of $r_{\rm h}\propto M^{1/3}$ (resulting in constant densities) for masses above which the maximum volume density is reached. Depending on what the threshold density would be, this could limit the survival of massive clusters during the rapid-disruption phase since $t_{\rm cce}\propto\rho_{\rm h}$.
\item
The simple model presented here is fundamentally unable to address the spatial evolution of cluster populations beyond the simple classification of clusters residing in the host galaxy disc or halo. How would GC populations evolve spatially under the conditions described in the proposed model? While this question should ideally be addressed using numerical simulations, it is possible to get a rough idea by considering the current literature. There exists a wide range of models in which the GCs are initialised in the galaxy halo (i.e.~at $t_{\rm merge}$ in our model, see Figure~\ref{fig:tschem}) and lose mass (largely) independently of the galactic environment \citep[e.g.][]{chernoff90,mclaughlin08,prieto08,li14}. Such models can reproduce the observed GCMF, colour bimodality, and the spatial distribution of GCs \citep[e.g.][]{prieto08,li14}. Although the cluster disruption model used here is environmentally dependent in general, it is most strongly so during the rapid-disruption phase in the host galaxy disc. Given that very few GCs are destroyed during the halo phase, they should be (re)distributed during hierarchical galaxy growth in the same way as in environmentally-independent models. This suggests that the two-phase GC formation model should also yield spatial distributions at $z=0$ that are consistent with those observed.
\item
Thus far, we have considered massive cluster formation at a particular redshift $z=3$ taken to be representative of the formation redshift of present-day GCs. How do the model predictions change when accounting for a realistic spread of GC formation redshifts and progenitor galaxy metallicities? Or conversely, what is the exact redshift range over which most of the GCs formed in our model? Some models have required GC formation to take place at $z=7$--$15$, i.e.~near the epoch of reionization, based on their present-day spatial distribution and hence their (initial) binding energies \citep[e.g.][]{moore06,brodie06}. However, these models did not account for the two-phase formation and disruption physics presented here and therefore they have limited information on the binding energies at which (the surviving) GCs actually formed. Including the appropriate physics should change the inferred range of GC formation redshifts.
\item
A related question is how the formation redshift range varies with the host galaxy mass or GC metallicity. Many models assume that metal-poor GCs formed in low-mass galaxies before the metal-rich ones did in massive galaxies \citep[e.g.][]{muratov10}. A similar relation is suggested by relative age measurements of GCs \citep[e.g.][]{forbes10}. However, the star formation histories of low-mass galaxies peak later in cosmic time than those of massive galaxies \citep[often referred to as `downsizing', e.g.][]{cowie96}. How can these two observations be united? Did metal-poor GCs indeed form at higher redshift than metal-rich GCs? Or do they trace the peak star formation activity of their host galaxies? Because the merger (and hence GC `liberation') rate increases with redshift, the peak of the GC formation history should be shifted towards higher redshifts than the peak star formation history of the host galaxy. Does this shift depend on the galaxy mass and does it therefore affect the relation between the GC metallicity and formation redshift?
\item
What is the reason for the existence of chemically enriched stellar populations in GCs? All present GC enrichment models do not rely on special, early-Universe physics, implying that the responsible process should also take place in nearby YMCs. Because YMCs are seen to be devoid of gas after a few~$\myr$, no evidence has yet been found for the multiple episodes of star formation required by the AGB and FRMS enrichment models. The search for ongoing or episodic star formation in YMCs should continue in order to test these scenarios. At the same time, an effort should be made to find evidence for the sweeping up of stellar ejecta by the protoplanetary discs of pre-main sequence stars, which is a critical test for the EDA enrichment model. All of these tests can be made observationally at high spatial resolution in the local Universe. In addition, the evidence is mounting that none of the existing enrichment models can explain the observed abundance patterns \citep{bastian15}, highlighting a clear need for new models, which (contrary to the previous generation) are not exclusively motivated by stellar evolution.
\item
Even if the majority of GCs has chemical abundance patterns consistent with a short formation time-scale, a handful of Galactic GCs show a spread or bimodality in their Fe abundance (e.g. $\omega$Cen). Some subset of these GCs likely did not form as regular YMCs, but may be the remnants of dwarf galaxy nuclei \citep{lee09}. Given this example, which fraction of GCs formed through mechanisms other than regular YMC formation? Answering this question requires current GC formation theories to be extended such that the numbers of these `alternative' GCs in present-day galaxy haloes can be predicted. For instance, coupling small-scale nuclear cluster formation models to the galaxy accretion histories of massive haloes from the Millennium Simulation enables their contribution to present-day GC populations to be quantified. A recent example of such an analysis was presented by \citet{pfeffer14}, who find that at most a handful of Galactic GCs may be stripped dwarf galaxy nuclei.
\item
In spite of all the evidence discussed in this work, it could be possible that for some reason the GCs observed today are not the products of regular cluster formation at $z>2$. However, we have also seen that the products of regular YMC formation in the high-redshift Universe should have survived until the present day, and that they should have properties very similar to those of the GC populations seen in nearby galaxies. If GCs formed differently, the question arises what happened to large numbers of YMCs that must have formed across cosmic time -- in the high-pressure environments of $z>2$ galaxies, a single YMC should have been formed for every $\sim10^6~\msun$ in stars. While our two-phase model requires a large fraction of these to have been destroyed, it seems unlikely that they could have disappeared altogether, as no known mechanism exists for disrupting a cluster with a mass of $10^6~\msun$ and a radius of $1.5~\pc$ within a Hubble time. Even in the $\feh=-0.6$ galaxy model, which is the most disruptive environment considered here, the short dynamical time of such a cluster allows it to respond adiabatically to tidal perturbations, leading to a disruption time-scale of $\sim10^2~\gyr$. Next to having to reproduce the observed properties of nearby GC populations, alternative GC formation theories would therefore also face the challenge of explaining why no products of regular YMC formation in $z>2$ galaxies would have survived until the present day.
\end{enumerate}

The above list of open questions represents a relatively modest sample from the wide range of unsolved problems in the field of GC formation and evolution. Rather than being exhaustive, it is intended to highlight a number of areas that would particularly benefit from a concentrated effort in the next several years. As indicated above, many of these questions can be addressed using the new observational facilities that should come online in the near future (e.g.~JWST) or have recently started operations (e.g.~ALMA). In terms of theoretical or numerical work, several of the above questions can be addressed by considering specific physical mechanisms in isolation, whereas others require a more holistic approach. In order to test GC formation theories, it will be essential to develop a direct connection between GC formation physics and the observable properties of the GC population at $z=0$. For instance, the two-phase GC formation model presented here is highly suitable for implementation in semi-analytic models of galaxy formation, as well as for verification and improvement using numerical simulations of galaxy formation. Above all, the current state of the field unequivocally shows that GC formation is a multi-scale, multi-physics problem, that is being constrained from a wide range of different approaches, both in terms of the physics considered and the methods used. The only way of turning the physics of GC formation into a quantitative and predictive science is to keep combining these different insights.

\section{Summary and conclusions} \label{sec:concl}
The purpose of this work is to address the question whether the products of young massive cluster (YMC, $M>10^4~\msun$) formation in the high-redshift Universe may have survived until the present day, and if these relics would be consistent with the properties of the globular cluster (GC) populations seen in nearby galaxies. In \S\ref{sec:constr}, we summarised the state of the literature related to this question. Using the observational (\S\ref{sec:obs}) and theoretical (\S\ref{sec:phys}) constraints on GC formation from nearby YMC formation, star formation in $z>2$ galaxies, and present-day GC populations, we derived a simple but quantitative, two-phase `{\it shaken, then stirred}' model for GC formation (\S\ref{sec:model}). Its main conclusions are as follows.
\begin{enumerate}
\item
The properties of GCs are consistent with being the products of the regular star and cluster formation process in the high-pressure environments of $z>2$ galaxies, i.e.~at the epoch of the peak cosmic SFR density. The early evolution of these clusters was characterised by rapid tidal disruption through encounters with dense gas clouds within the gas-rich host galaxy discs. However, the clusters were saved from tidal destruction in their dense natal environments by the redistribution of matter during galaxy mergers and/or tidal stripping, which allowed them to be `liberated' into their host galaxy haloes, leading to their spatial distribution as observed today and marking the onset of an episode characterised by slow tidal disruption.
\item
This simple model reproduces a wide range of observables at $z=0$ across a broad range of galaxies (\S\ref{sec:comp}): (1) the mass spectrum of GCs, (2) the relation between the number of GCs per unit stellar mass and the metallicity or galaxy stellar mass, (3) the colour and metallicity bimodality, and (4) the constant ratio between the GC population mass and the host galaxy's dark matter halo mass. This variety of successes of the model is compelling, as it shows that combining the physics of star cluster formation and hierarchical galaxy formation naturally leads to the GC populations observed at $z=0$.
\item
We find that including the rapid-disruption phase in GC formation models is essential to avoid making several ad-hoc assumptions that would be necessary when only capturing the evolution of GCs in galaxy haloes. In halo-only models, the relations between the number of GCs and their host galaxies are attained after an entire Hubble time of GC evolution by evaporation. By contrast, the new model predicts that the relations between the number of GCs and their host galaxies were already in place at $z=1$--$2$, whereas (primarily) the radii (and to a limited degree the masses) of GCs were shaped by their quiescent evolution in galaxy haloes.
\item
The high masses of GCs are a remnant of the Toomre mass during their formation, requiring surface densities of $\Sigma=100$--$300~\msun~\pc^{-2}$ and volume densities of $\rho_{\rm ISM}=1$--$5~\msun~\pc^{-3}$, which were common conditions in galaxies at $z>2$. Due to the disruptive, high-pressure environments in these galaxies, only massive clusters could survive until the time of the first galaxy merger and the corresponding migration of clusters into the galaxy halo (i.e.~when the rapid-disruption phase ended and the clusters became GCs according to the definition of \S\ref{sec:intro}). For this reason, {\it GC are typically massive}.
\item
As the cosmic density decreases with cosmic time, so do both the Toomre mass and the galaxy merger rate. As a result, at lower redshifts fewer new-born clusters can survive long enough to escape into galaxy haloes, making it highly unlikely for them to survive over cosmic time-scales. For this reason, {\it GCs are typically old}. Because the metallicity increases with cosmic time and galaxy mass, the peaking galaxy merger and GC survival rates at high redshift and low galaxy masses also imply that {\it GCs are typically metal-poor}.
\item
During the hierarchical growth of galaxies, the native GC population remains associated with the host galaxy spheroid, whereas accreted GCs from stripped satellite galaxies mainly reside in the halo. Because even the native GCs must have escaped the star-forming component of the host galaxy to survive over long time-scales, the typical galactocentric radii occupied by GCs extend further than those of field stars, showing that {\it a substantial fraction of the GC population resides in the halo}.
\item
All models for the formation of an enriched population of stars in GCs predict that enrichment requires a minimum cluster mass (or density). By comparing the enriched stellar mass in GCs and in the field to the mass loss histories predicted by our model, we find that this minimum mass-scale is $M\sim10^{5.5}~\msun$. If true, the high initial masses of GCs ($M_{\rm i}\ga10^5~\msun$, even for a current GC mass of $M\sim10^2~\msun$) ensure that the minimum mass criterion is generally satisfied and hence {\it the majority of GCs should display the observed multiple populations}.
\end{enumerate}
In \S\ref{sec:intro}, we defined a GC as ``a gravitationally-bound, stellar cluster that in terms of its position and velocity vectors does not coincide with the presently star-forming component of its host galaxy''. This definition excludes clusters occupying the same part of position-velocity space as the host galaxy's gaseous component. It thereby minimises GC destruction and thus represents a definition analogous to an `anthropic principle' for long-lived stellar clusters. We preferred this definition over previous ones adopted in the literature, based on metallicity (`metal-poor'), mass ($M=10^4$--$10^6~\msun$), age ($\tau\sim10^{10}~\yr$), location (`in the halo'), or chemical abundance variations (`multiple stellar populations'). The results summarised above show that all of these criteria are implied by combining our definition with the physics of GC formation.

In the presented model, the `universal' properties of GC populations expressed by the above range of criteria emerges irrespectively of their widely-varying present-day environments because the conditions of their formation were so similar. However, contrary to previous GC formation theories, our model also allows exceptions to these criteria, e.g.~due to the stochastic nature of galaxy mergers, which in individual cases will occur on time-scales shorter or longer than the cosmic average. For instance, if a merger takes place immediately after a burst of massive cluster formation, the clusters are liberated into the halo shortly after their birth, giving rise to (possibly metal-rich) lower-mass, young GCs.

The two-phase GC formation model explains a substantial number of observed characteristics of GCs (\S\ref{sec:comp}) and has a broad range of associated predictions and implications (\S\ref{sec:pred}), e.g.~for constraining the physical processes involved in GC formation (\S\ref{sec:discdis}--\ref{sec:df}), as well as for testing the leading models for multiple stellar populations in GCs (\S\ref{sec:mi}--\ref{sec:acc}). For instance, it is found that GCs were typically only 2--3 times more massive at birth and that most GCs could not have survived the high gas densities required for pristine gas accretion. These findings challenge the AGB and FRMS enrichment models.

Next to these insights gained from the model, several open questions remain. These have been discussed in \S\ref{sec:open}, but one particular question is reiterated here. The evidence discussed in this work suggests that the products of regular YMC formation in the high-redshift Universe should have survived until the present day, and that they should have properties very similar to those of the GC populations seen in nearby galaxies. This conclusion lends support to the idea that YMCs in nearby galaxies can be used to test GC formation models. If GCs would not be the products of regular cluster formation at $z>2$, the field would be facing a new problem: given the enormous number of YMCs produced in the history of the Universe and their known longevity, where did they go?

The ideal cosmic conditions for GC formation existed at $z>2$ due to the high mass-scales for gravitational instability and the frequent galaxy mergers that redistributed GCs into the silent waters of galaxy haloes. The likelihood of GC formation has since declined, but GCs can still form in the local Universe. When young, massive clusters are liberated into their host galaxy halo by a merger, their long-term survival has been secured, effectively turning them into GCs. Although the GC population does erode with time, it presently does so at a much lower rate than in the past -- the cluster disruption rate peaked in the host galaxy discs where GCs formed. Once thought to be the relics of a bygone era, the evidence discussed in this work suggests that GC populations will survive, and in rare cases even grow, for at least another Hubble time.

\section*{Acknowledgements}
JMDK is greatly indebted to Nate Bastian, Jim Dale, Bruce Elmegreen, Mark Gieles, Henny Lamers, Chervin Laporte, S\o ren Larsen, Steve Longmore, Simon Portegies Zwart, and Marina Rejkuba for several years of discussions from which large parts of this work were drawn. In addition, Nate Bastian, Bill Harris, Gretchen Harris, and Steve Longmore are gratefully acknowledged for their careful reading of the manuscript. JMDK thanks Angela Adamo, Shy Genel, Michael Hilker, Phil Hopkins, Mark Krumholz, Claudia Lagos, Ben Moster, Thorsten Naab, Fran\c{c}ois Schweizer, Linda Tacconi, and Simon White for helpful conversations and suggestions. The anonymous referee is thanked for constructive comments that improved the presentation of this work.

\bibliographystyle{mn2e}
\bibliography{mybib}

\label{lastpage}
\bsp

\end{document}